%% file: main.tex
\documentclass[twocolumn]{aastex701}

\defcitealias{Biswas2026}{Biswas26}
\defcitealias{Bisigello2025}{Bisigello25}
\defcitealias{Schreiber2015}{Schreiber15}
\defcitealias{Barrufet2023}{Barrufet23}
\defcitealias{Cutler2026}{Cutler26}
\defcitealias{Muzzin2025}{Muzzin25}
\defcitealias{Barrufet2025}{Barrufet25}
\defcitealias{Barrufet2026}{Barrufet26}
\defcitealias{Popesso2023}{Popesso23}
\defcitealias{PerezGonzalez2023}{P\'erez-Gonz\'alez23}
\defcitealias{Zavala2021}{Zavala21}
\makeatletter
\newcommand{\bibalias}[1]{\@ifundefined{al@#1}{}{ (\csname al@#1\endcsname)}}
\makeatother

\begin{document}

\title{Illuminating Hubble-Dark Galaxies at Cosmic Noon and Beyond with JWST/MINERVA UDS-DR0}

\author[0009-0009-8386-6232]{Advay Almal}
\affiliation{Department of Physics, Ashoka University, Sonipat, 131029, Haryana, India}
\email{advay.almal@outlook.com}

\author[0009-0002-2173-0953]{Pralay Biswas}
\affiliation{National Centre for Radio Astrophysics, Tata Institute of Fundamental Research, Pune 411007, India}
\email{pbiswas@ncra.tifr.res.in}

\author[0000-0002-1345-7371]{Yogesh Wadadekar}
\affiliation{National Centre for Radio Astrophysics, Tata Institute of Fundamental Research, Pune 411007, India}
\email{yogesh@ncra.tifr.res.in}

\begin{abstract}

At $z > 3$, rest-optical starlight redshifts beyond the HST's wavelength coverage, imposing a strong rest-UV-bright bias on legacy galaxy surveys. The JWST/NIRCam is revealing a growing population of previously unseen, rest-UV-faint sources called HST-Dark galaxies, extending up to and beyond the epoch of reionization, challenging conventional empirical models and statistics of galaxy populations. Using UDS field observations from JWST/MINERVA,  we characterize the largest reported sample of HST-Dark galaxies. We employ multi-band photometry in 8 wide-bands, $\leq 8$ medium-bands, 4 JWST/MIRI filters, and ancillary HST imaging to select 372 HST-Dark sources, model their spectral energy distributions using BAGPIPES and analyze their morphology using \texttt{statmorph}. Above a completeness limit ($z \geq 2$), these galaxies have high redshifts (median $z \sim 4.15$), high stellar masses (median $M_\star\sim 10^9 M_\odot$), moderate star-formation rates (median SFR $\sim 22 \,M_\odot \mathrm{yr^{-1}}$), and very high dust attenuation (median $A_V$ $\sim$ 1.85), with more compact sizes than equal-mass non-HST-Dark counterparts, indicating higher baryonic densities. Below the completeness limit ($z < 2$), the remaining galaxies show properties characteristic of High-Extinction Low-Mass (HELM) sources. The high-z sample demonstrates a $\sim 20-30\%$ previously-missed contribution to the redshift-binned, dust-unobscured integrated cosmic star formation rate density (CSFRD) at $z \gtrsim 5$. Our results indicate that early-Universe cosmological trends, such as the evolution of the CSFRD, may require recalibration to account for these previously missed sources, motivating deeper, wider JWST observations to better constrain this population and provide a more complete census of galaxies across cosmic time.

\end{abstract}


\section{Introduction}
\label{sec:introduction}
\input{introduction}

\section{Data}
\label{sec:data}
\input{data}

\section{Morphological Analysis}
\label{sec:morphology}
\input{morphology}

\section{Sample Selection}
\label{sec:sample-selection}
\input{sample_selection}

\section{Results and Discussion}
\label{sec:results}

\input{results}

\section{Summary and Conclusions}
\label{sec:conclusion}
\input{conclusion}

\begin{acknowledgments}
We thank the PRIMER and MINERVA collaboration teams for designing the observing programs, developing the data reduction pipelines, and making high-level data products publicly accessible to the astronomical community. We obtained the data used in this work from the MINERVA survey website at \url{https://minerva.colorado.edu/}. We also acknowledge the longstanding community effort to observe the CANDELS and 3D-HST extragalactic deep fields.
This work is based on observations made with the NASA/ESA/CSA James Webb Space Telescope and the NASA/ESA Hubble Space Telescope. The data were obtained from the Mikulski Archive for Space Telescopes (MAST) at the Space Telescope Science Institute, which is operated by the Association of Universities for Research in Astronomy, Inc., under NASA contract NAS 5-03127 for JWST and NAS 5-26555 for HST. These observations are primarily associated with JWST Cycle 1 Treasury program 1837 (PRIMER) and Cycle 4 Treasury program 7814 (MINERVA) in the Ultra-Deep Survey field, along with supporting JWST observations from the JWST-GO-6368, JWST-GO-2565, JWST-GO-8559 and JWST-GO-5224 programs and archival HST observations from the CANDELS (HST-GO-12064, and supporting program HST-GO-16872) and 3D-HST (HST-GO-12177 and HST-GO-12328) programs.
Some of the data products presented herein were retrieved from the Dawn JWST Archive (DJA). DJA is an initiative of the Cosmic Dawn Center (DAWN), which is funded by the Danish National Research Foundation under grant DNRF140.
AA acknowledges support from Ashoka University. AA thanks NCRA for support during the Visiting Students' Research Programme. YW and PB acknowledge the support of the Department of Atomic Energy, Government of India, under the project title ``Next Generation Instrumentation for Radio Astronomy'', UID number RTI4017. We thank Sam Cutler for useful discussions and for readily sharing some MINERVA data products through private communication, ahead of public release. We thank Rashmi Gottumukkala for useful discussions, especially concerning the Star-Forming Main-Sequence. AA expresses gratitude to Prof. Somak Raychaudhury, Prof. Kandaswamy Subramanian, and, in particular, Prof. Dipankar Bhattacharya at Ashoka University for their helpful discussions and academic concessions that made this work possible. AA thanks T. Agarwal, S. Gupta and S. Prakash for guidance and support in carrying out this inaugural work.

\end{acknowledgments}

\section{Data Availability}
\label{sec:data-availability}

The MINERVA UDS-DR0 mosaics and SUPER photometric catalog used in this work are publicly available from the MINERVA data release \citep{Muzzin2025, Cutler2026}, and the NIRSpec spectroscopic redshifts are available from the DAWN JWST Archive. 

The results of our analysis include our HST-Dark science catalog, with the BAGPIPES outputs and \texttt{statmorph} measurements for all 372 galaxies in our sample. The catalog is available in the online version of the paper as supplementary material.

\facilities{HST (ACS, WFC3), JWST (NIRCam, MIRI, NIRSpec)}

\software{astropy \citep{Astropy2013, Astropy2018, Astropy2022}, BAGPIPES \citep{Carnall2018}, Cloudy \citep{Ferland2017}, FITSMap \citep{Hausen2022}, MultiNest \citep{Feroz2008, Feroz2009, Feroz2019}, statmorph \citep{RodriguezGomez2019}, TOPCAT \citep{Taylor2005}, matplotlib \citep{matplotlib}, pandas \citep{pandas2010, pandas2020}, numpy \citep{numpy}}

\bibliographystyle{aasjournalv7}
\bibliography{references}

\end{document}

%% file: introduction.tex
Understanding the formation and evolution of galaxies and, by extension, the large scale structure of the Universe, invariably requires tracing the change in their properties across cosmic time. Constructing unbiased, representative samples of galaxies is the first step in this direction.

The telescopes and instruments used to collect astronomical data impose systematic constraints on the properties of the galaxy populations under study. The Hubble Space Telescope (HST) has transformed our understanding of the distant Universe, with its remarkable sensitivity and ability to resolve rest-frame optical and ultraviolet (UV) starlight from galaxies at $z < 3$. Observations in other bands have often been follow-up or correlative studies of HST-detected sources. Consequently, the sensitivity and wavelength coverage of the HST have shaped our understanding of mass build-up, chemical enrichment, the galaxy main sequence, the cosmic star formation rate density (CSFRD), and other empirical relations across cosmic time. At higher redshifts, HST probes only rest-frame UV emission, so legacy observations of high-redshift sources such as Lyman-break galaxies carry a systematic UV bias that disproportionately affects our attempts at a census of galaxy populations at these epochs. 

New facilities allowing us to push past these limitations are now revealing a previously ``dark'' population that legacy HST surveys have systematically missed. With its superior resolution, sensitivity, and broad near-infrared (NIR) wavelength coverage, the James Webb Space Telescope (JWST) is a leading driver of the study of these ``HST-Dark galaxies''.

\begin{figure}[h!]
    \centering
    \includegraphics[width=\columnwidth]{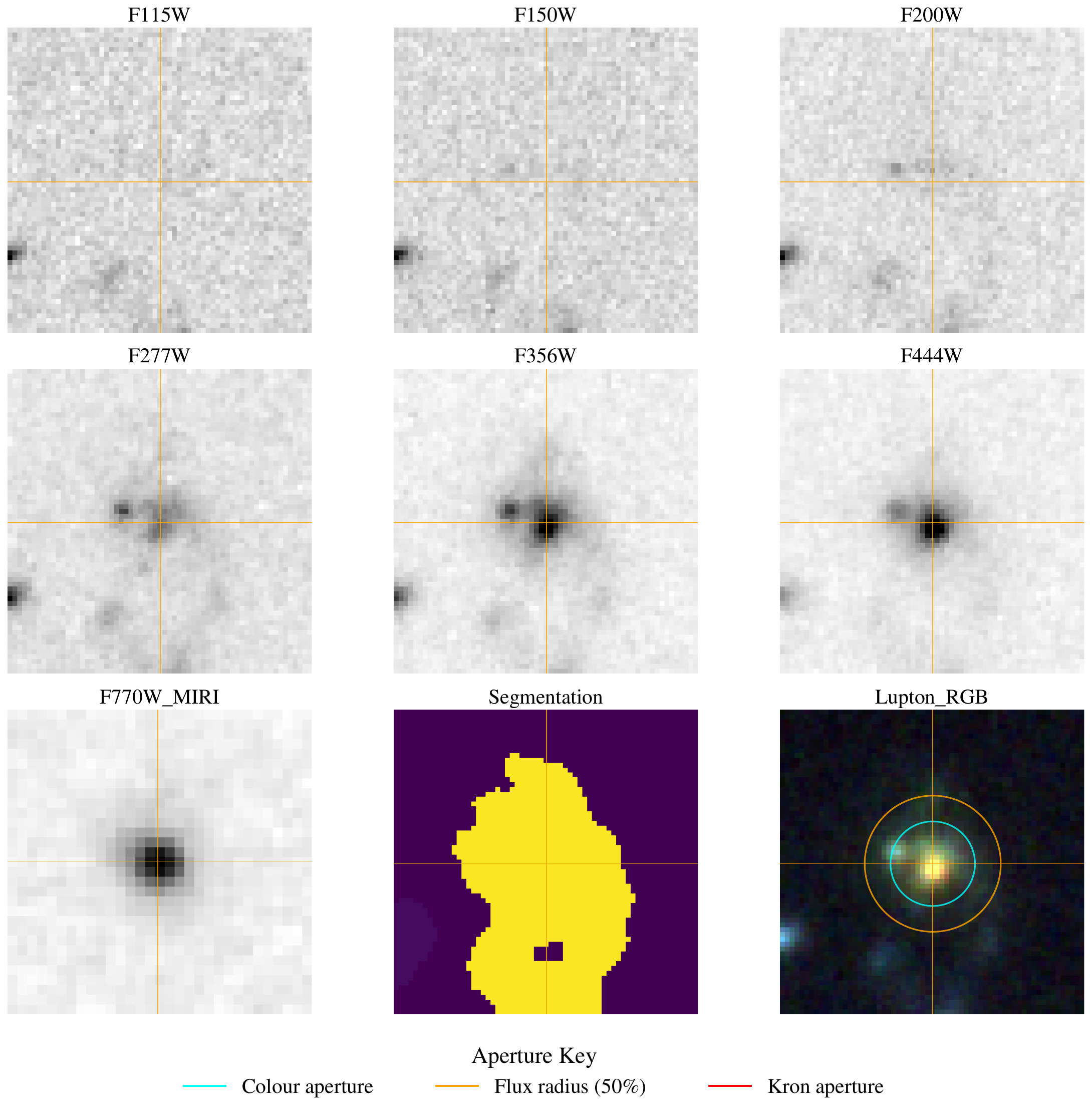}
    \caption{An HST-Dark galaxy [\textit{MINERVA UDS-DR0 ID} (hereafter, \textit{MinID) $\# 1018752$}] imaged in 7 NIRCam broad band filters and one MIRI filter, with the segmentation map and photometric apertures plotted. The galaxy shows the characteristic HST-Dark behavior of increasing intensity with increasing wavelength. F150W ($1.5\,\mu$m) approximates the long-wavelength limit of HST/WFC3. The source is undetected in F150W, while a large, morphologically interesting source appears at longer NIR wavelengths.}
    \label{fig:Mosaic_2_1018752}
\end{figure}

Because the detection limits of HST have long been known, several studies have attempted to study these hidden HST-Dark populations using ALMA and Spitzer. Sub-millimeter galaxies (SMGs), detected by ALMA, are typically extreme starbursts with $\mathrm{SFR} \sim 1000\,M_\odot\,\mathrm{yr}^{-1}$ at $z > 4$. Their very low sky density ($\sim 0.01$ arcmin$^{-2}$) makes them negligible contributors to the integrated CSFRD and other such number- or volume-density-dependent quantities \citep{Riechers2013, Marrone2018}. They are heavily dust-attenuated, with a median $A_V \sim 2.10$ \citep{daCunha2015}, and lie at the high-mass, high-SFR end of the $z = 4$ main sequence presented by \citet[][hereafter \citetalias{Schreiber2015}]{Schreiber2015}. A small population of massive quiescent galaxies has also been observed up to $z \sim 3-$5; due to their rarity in legacy observations at these epochs, they make little contribution to global quantities such as the cosmic baryonic mass density when inferred from pre-JWST data \citep{Tanaka2019, Carnall2020, Santucci2020, Valentino2020, Leung2026, Stevenson2026}.

Another previously studied HST-Dark population comprises Spitzer-selected HST-Dark galaxies \citep{Huang2011, Caputi2015, Stefanon2015, Wang2019b, Franco2018, Alcalde2019, Fudamoto2021}. These sources are faint or undetected in deep HST imaging but are detected in the longer-wavelength Spitzer/IRAC bands, usually selected by a color criterion between the HST $H$ band and the Spitzer [4.5] band (4.5\,$\mu$m), and subsequently followed up with ALMA \citep[e.g.,][]{Wang2019b}. They are also bright, and hence massive and highly star-forming, but they occupy a less extreme region of parameter space compared to the SMGs. Their properties are poorly constrained due to Spitzer's lower sensitivity and poorer resolution than HST, with the broad PSF causing significant blending and source confusion in crowded deep fields, and the limited number of IRAC photometric bands introducing systematics and errors in SED modeling.

Together, these studies reveal a physically extreme, low-density population detected only by pushing legacy instruments to their limits. Their sub-dominant contribution to number/volume-density dependent cosmological quantities suggests that the dominant contributors to early stellar mass build-up and cosmic star formation were likely less extreme sources that are possibly missing from pre-JWST surveys.

JWST/NIRCam \citep{Rieke2003, Rieke2005, Greene2017, Gardner2023, Rieke2023}, with its broad multiband coverage ($\lambda \sim 1$--5\,$\mu$m), very high spatial resolution and unmatched sensitivity, allows us to push our detection limits significantly and to identify, resolve, and analyze many more such red, optically dark sources.

The Cosmic Evolution Early Release Science (CEERS) survey \citep{Finkelstein2025} was one of the first extragalactic observational programs with the JWST. \citet[][hereafter \citetalias{Barrufet2023}]{Barrufet2023} studied 30 HST-Dark sources selected by the JWST/NIRCam colors in this pioneering survey. These dark galaxies were found to have lower stellar masses and SFRs than the SMGs and Spitzer-selected galaxies, and were very dusty systems lying on or near the star-forming main sequence at $z \sim 2$--8, with a sky surface density of $\sim 0.8$ arcmin$^{-2}$. \citetalias{Barrufet2023} modeled their spectral energy distributions (SEDs) with the BAGPIPES software \citep{Carnall2018} using 12-band photometry, including ancillary HST imaging. The small sample size leaves the derived trends and properties poorly constrained, serving as indicative results that motivate deeper and wider surveys to test for the effect of cosmic variance and to uncover the less extreme part of the population.

\citet[][hereafter \citetalias{Biswas2026}]{Biswas2026} used 27-band photometry from the UNCOVER/MegaScience SUPER Catalog \citep{Bezanson2024, Suess2024} \footnote{UNCOVER/MegaScience SUPER Catalog: \url{https://jwst-uncover.github.io/\#releases}}, containing observations of the Abell 2744 lensing cluster, to identify and study 207 HST-Dark galaxies, 113 of which lie at $z > 2$. These sources form the much fainter, low-mass end of the HST-Dark population and are detected due to lensing magnification by Abell 2744 and the deep imaging depths in UNCOVER/MegaScience. They span a broad range in stellar mass and SFR, with $M_\star$ as low as $\sim 10^{7.5}\,M_\odot$, and are moderately dusty galaxies on the main sequence. The $z < 2$ population is extremely low-mass and very dusty, consistent with the High-Extinction Low-Mass (HELM) population defined by \citet[][hereafter \citetalias{Bisigello2025}]{Bisigello2025}. 

Neither CEERS nor UNCOVER/MegaScience covers a wide area. Moreover, neither of these studies samples the typical (median) component of the population in a statistically significant manner. CEERS does target this component, but with only 30 sources in a small part of the sky, it serves as a diagnostic study with large Poisson errors on the derived quantities. Cosmic variance and sampling errors may therefore bias the derived parameters. A much larger sample from a significantly wider survey is required to corroborate existing results and constrain the properties of the population with greater statistical significance.

The MINERVA survey \citep[][hereafter \citetalias{Muzzin2025}]{Muzzin2025} will address this need. As the widest-area JWST/NIRCam medium-band survey to date, it helps mitigate cosmic variance and, combined with other JWST and HST programs, provides photometry in 26--35 filters, thus providing robust photometric coverage across the optical/NIR regime. Of its planned 542 arcmin$^2$ of NIRCam imaging, $\sim 234$ arcmin$^2$ has been released so far and is used in this paper; this is already 4--5 times wider than UNCOVER/MegaScience, and after completion, this factor will rise to $>10$. The medium-band and MIRI filters improve the quality of the fits and break degeneracies in the modeling \citep{Suess2024}. MINERVA reaches depths comparable to CEERS and should therefore detect a similar, median HST-Dark population in greater numbers. This allows us to rigorously test the observed trends and further constrain the physical properties of these optically invisible galaxies up to cosmic noon and beyond.

Robust and complete demographic studies of galaxies require multi-color data for photometric redshift estimation and SED modeling. Broad-band-only photometry limits the accuracy of SED modeling, primarily because of the strong emission lines common in high-redshift spectra that cannot be well constrained by wide-bands, leading to photometric degeneracies, and the likely confusion between the Balmer and Lyman breaks due to coarse SED sampling. For quiescent galaxies with dusty star formation and prominent absorption complexes, broad-band-only photometry can misidentify apparent ``breaks'' as nonexistent Balmer/4000\,\AA\ breaks, leading to further degeneracies between inferred properties. Such degeneracies prevent robust estimates of rest-frame colors, stellar masses, redshifts, and other parameters, and also affected Spitzer studies well before JWST \citep[e.g.,][]{Labbe2013, Stark2013, Smit2014}.  

Medium-band photometry samples well-known emission lines and finer spectral features while remaining more observationally efficient than narrow bands, and some JWST studies demonstrate that medium-band data improve photometric redshift and stellar mass estimates \citep{Sarrouh2024, Suess2024}. NIRCam medium bands resolve degenerate Balmer breaks in distant galaxies (e.g., \citealt{Trussler2025, Mintz2025, Khullar2026}), dusty galaxy SEDs \citep{Martis2025, Lorenz2025} and strong line emitters (e.g., \citealt{Withers2023, Rinaldi2023, Williams2023, Wold2025}; Porraz et al., in prep.). However, existing JWST surveys with good medium-band coverage are limited by cosmic variance and volume.

MINERVA adds the missing medium bands to treasury JWST broad-band data, addressing the systematic failures of broadband-only SED fitting. In this paper, we analyze the publicly released MINERVA DR0 data \citep[][hereafter \citetalias{Cutler2026}]{Cutler2026} containing  PRIMER (GO 1837, PI: J. S. Dunlop; \citealp{Donnan2024})  and MINERVA observations of the UDS field--the largest field by area of the 4 CANDELS fields with planned MINERVA observations.

A galaxy can be HST-Dark because it is too faint or too red--or, most often, both-- given the HST sensitivity and wavelength coverage. This dual criterion broadens the parameter space that these sources can occupy and increases the variance in the properties of the population. In this paper, we analyze the galaxies in the four-dimensional parameter space of redshift $z$, V-band dust attenuation $A_V$, star formation rate (SFR), and stellar mass $M_\star$. Specific combinations of these parameters make a source HST-Dark yet reliably detectable at our JWST survey limits. Increasing $z$ or $A_V$ makes a galaxy redder, because the UV/optical light is either redshifted out of HST-coverage or absorbed by dust, respectively. Low SFR and low stellar mass reduce the overall luminosity, making the galaxy fainter. These are the typical properties of an HST-Dark source, although a departure in any one parameter can be compensated for by opposite changes in the others.

The outline of this paper is as follows: in Section \ref{sec:data}, we describe the imaging data and photometry used by us as inputs, including a brief summary of the modeling pipeline in Section \ref{sec:bagpipes}. Section \ref{sec:morphology} describes the non-parametric morphological measurements we performed using the \texttt{statmorph} package, and the pipeline we wrote to do so. Section \ref{sec:sample-selection} details the sample selection procedure and the justifications for each round of cuts. Section \ref{sec:results} analyzes the properties of our final sample and discusses the physical implications of the inferred correlations and derived statistics. Finally, Section \ref{sec:conclusion} summarizes the methodology and inferences from our work, while highlighting important avenues for follow-up investigation. 

Throughout this paper, we adopt the AB magnitude system as presented by \citet{Oke1983}  and the Planck18 cosmology \citep[][Table 2; TT, TE, EE + lowE + lensing + BAO]{Planck2020}, with $H_0 = 67.7$ km s$^{-1}$ Mpc$^{-1}$ and $\Omega_{\rm m} = 0.31$.

\begin{figure} [b]
    \centering
    \includegraphics[width=\columnwidth]{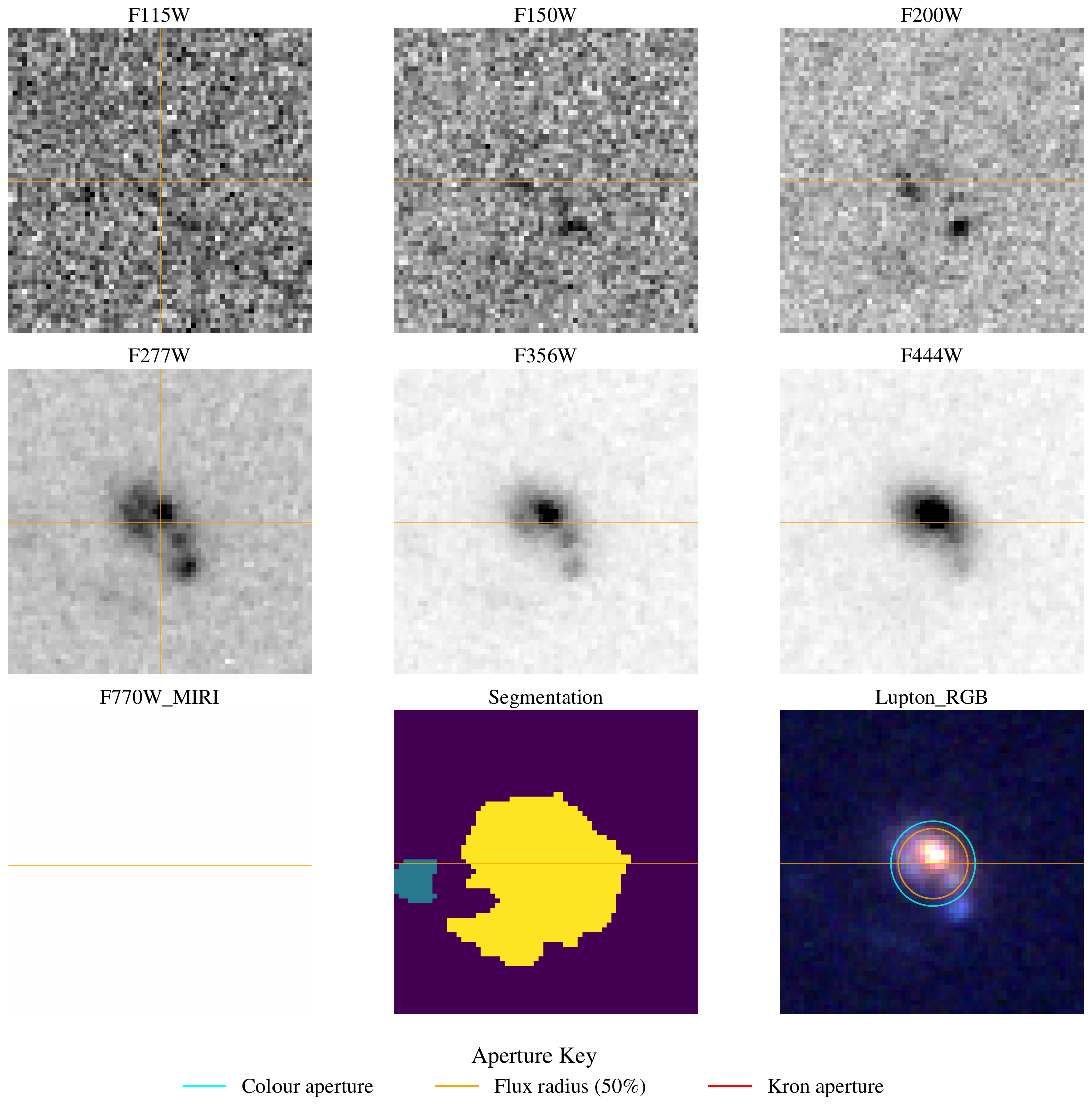}
    \caption{\textit{MinID \#1003714}, another demonstration of the typical HST-Dark pattern of increasing brightness with increasing wavelength. In the F150W (HST-limit) image, only the bright clumps in the bottom right quadrant are detected.}
    \label{fig:Mosaic_1_1003714}
\end{figure}

%% file: data.tex
The Medium Band Imaging with NIRCam to Explore ReVolutionary Astrophysics (MINERVA) survey \footnote{MINERVA website: \url{https://minerva.colorado.edu}} (PID 7814; PI: Muzzin; Co-PIs: Marchesini, Suess) is an ongoing JWST Cycle 4 treasury program using 8 NIRCam medium-band filters spanning 1--5\,$\mu$m (F140M, F162M, F182M, F210M, F250M, F300M, F360M, F460M), with MIRI \citep{Wright2023} parallels in F1280W and F1500W. It covers 4 of the 5 HST CANDELS fields (UDS, COSMOS, AEGIS and GOODS-N), which have extensive multiwavelength coverage from many telescopes, with a total of 542 arcmin$^2$ of NIRCam imaging and 275 arcmin$^2$ of MIRI imaging, substantially reducing both Poisson errors and cosmic variance. The planned NIRCam medium-band imaging area is $\sim$7 times larger than all existing extragalactic medium-band coverage, and among surveys with good medium-band coverage ($>14$ filters), MINERVA is shallow but the widest area survey.

In this paper, we analyze the MINERVA Data Release 0 containing data for the UDS field (UDS-DR0), the largest field by area observed by the MINERVA and PRIMER programs. HST/ACS F435W imaging from PID 16872 (PI: Grogin), with coverage similar to CANDELS, completes the coverage in the three primary HST optical bands.

The SPAM program (PID 8559, PI: K. Davis, Co-PI: R. Larsen) is a JWST Cycle-4 observing program that adds medium band data to the MINERVA footprint in the F335M, F430M \& F480M filters.

The Public Release IMaging for Extragalactic Research (PRIMER-UDS, e.g., \citealt[][]{Donnan2024}), a JWST Cycle 1 program with extensive NIRCam and MIRI coverage, consists of MIRI prime pointings in F770W and F1800W with NIRCam parallels in 7 wide bands (F090W, F115W, F150W, F200W, F277W, F356W, and F444W) and one medium band (F410M). The MIRI exposure time is $\sim$30 minutes per pixel per filter. 

MINERVA-UDS reuses the PRIMER APT footprint and position angle (for NIRCam medium band parallels), with different filters and a different dither and readout strategy, improving background subtraction and bad-pixel masking relative to PRIMER; the slightly longer observing time increases the MIRI integrations to $\sim$38 minutes per filter. The full MINERVA-UDS observations, with 54 MIRI prime pointings and corresponding NIRCam parallels, span 125 and 234 arcmin$^2$, respectively.

\begin{figure}[tb]
    \centering
    \includegraphics[width=\columnwidth]{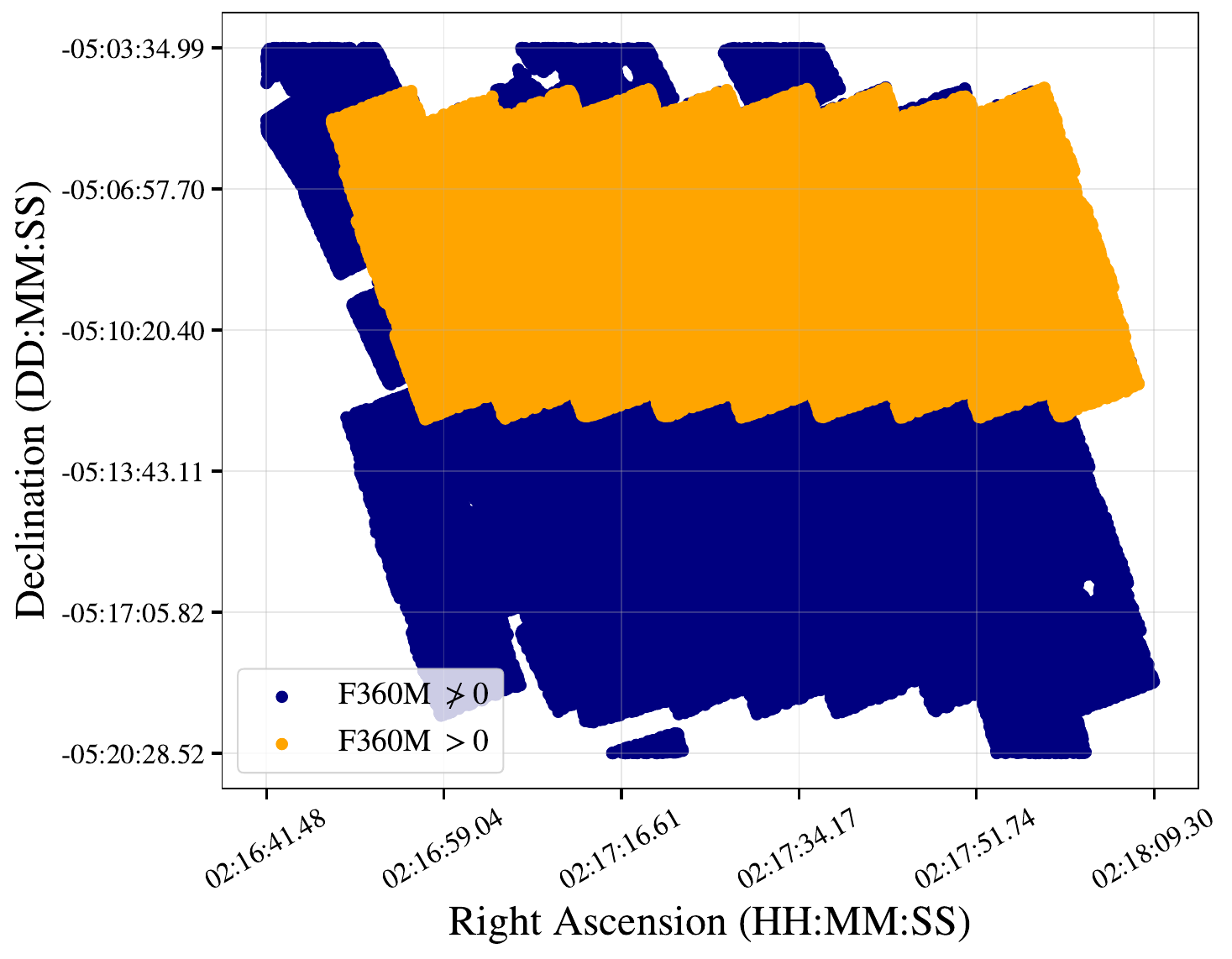}
    \caption{The MINERVA UDS-DR0 catalog in R.A./Dec. coordinates. Orange points show sources with MINERVA UDS-DR0 medium band coverage; navy blue points in the background show sources without this medium band data. F360M $>0$ is chosen as the criterion to depict the DR0 medium band coverage.}
    \label{fig:minerva-mband-coverage}
\end{figure}

The UDS-DR0 products used in this work contain all PRIMER data for the entire UDS field but only half of the MINERVA observations (the upper half of the field in R.A.--Dec. coordinates; Figure~\ref{fig:minerva-mband-coverage}). Data for the remaining part of the field have been collected, but not released publicly yet by the MINERVA team; we discuss the implications of this partial coverage in Section~\ref{sec:sample-selection}. In addition to this, MINERVA UDS-DR0 contains HST imaging in 6 filters from CANDELS and 3D-HST (F435W, F606W, and F814W from HST/ACS and F125W, F140W, and F160W from HST/WFC3/IR).

\citet{Muzzin2025} tested the MINERVA filter choice and exposure times with simulations, which show that the $2\times$ longer integration time of MINERVA relative to PRIMER reduces the $\sigma_{\mathrm{NMAD}}$ scatter in photometric redshifts and $\log_{10}(M/M_\odot)$ by factors of 3.7 and 1.78, respectively.

MINERVA-UDS has NIRSpec \citep{Jakobsen2022} spectroscopy from several major campaigns, including CAPERS (PID 6368, PI: Dickinson), EXCELS \citep{Carnall2024}, RUBIES \citep{deGraaff2025}, the NIRSpec GTO Wide Survey \citep{Maseda2024}, PID 2565 (PI: Glazebrook), and Mirage or Miracle (PID 5224, PI: Oesch \& Naidu). About 3100 high-confidence (class 3) NIRSpec spectroscopic redshifts are currently available in the DAWN JWST Archive \citep[DJA;][]{deGraaff2025, Heintz2025} for the MINERVA-UDS field.

\subsection{Photometry}
\label{sec:photometry}

All analysis in this paper is based on the MINERVA UDS-DR0 SUPER catalog \footnote{Link to MINERVA UDS-DR0 SUPER Photometric Catalog: \url{https://minerva.colorado.edu/data_release_products/uds/catalogs/MINERVA-UDSDR0.fits}} (\citetalias{Muzzin2025}; \citetalias{Cutler2026}), which includes PRIMER data for the full UDS field and MINERVA data for $\sim50\%$ of it; the remaining half of MINERVA-UDS has been observed and will appear in future data releases. The SUPER catalog measures photometry in the best aperture for each object using adaptive aperture selection with aperture diameters of 0.20$''$, 0.32$''$, 0.48$''$, 0.70$''$, 1.00$''$, and 1.40$''$. Below, we summarize the photometric pipeline and main features of the MINERVA UDS-DR0 SUPER catalog \citetalias{Cutler2026}, which is used as input for our SED modeling; further details are given in the README file \footnote{\url{https://minerva.colorado.edu/data_release_products/uds/catalogs/MINERVA-UDSDR0_README.txt}} on the MINERVA website.

To construct the SUPER catalog,  Source Extraction and Photometry (SEP) v1.4.1 \citep{Bertin1996, Barbary2016} was run on a noise-equalized Long Wavelength (LW) detection image (F277W+F356W+F444W), with aperture corrections following \citet{Whitaker2011}. Detected sources were cross-matched with previous JWST catalogs within 0.08$''$ and ID'd consistently.

Aperture photometry was performed in HST/ACS (F435W, F606W, F775W, F814W, F850LP), HST/WFC3 (F098M, F105W, F125W, F140W, F160W), and JWST/NIRCam (F090W, F115W, F140M, F150W, F162M, F182M, F200W, F210M, F250M, F277W, F300M, F335M, F356W, F360M, F410M, F430M, F444W, F460M, F480M) background-subtracted reduced images. All mosaics were measured at a uniform pixel scale of 0.04$''$/pixel, with the 0.02$''$/pixel Short Wavelength (SW) mosaics co-added to this scale. Fluxes reported in the catalog are in units of $F_\nu = 10$ nJy, corresponding to an AB magnitude zero-point of 28.9. The mosaics were processed with grizli \citep[v8.0;][]{Brammer2021} and are publicly available.

\begin{table*}[]
\begin{tabular}{llll}
\hline \hline
\textbf{Component}                      & \textbf{Parameter}        & \textbf{Prior Distribution}  & \textbf{Range/Value}        \\ \hline
Stellar Population                      & IMF                       & Fixed (Kroupa \& Boily 2002) & $-$                         \\ \hline
Star Formation History & Model                     & Delayed $\tau$               & $te^{-t/\tau}$              \\
                                        & Age ($Gyr$)               & Uniform                      & {[}0.01, Age of Universe{]} \\
                                        & $\tau$ ($Gyr$)            & Logarithmic                  & {[}0.01, 10{]}              \\
                                        & $\log(M_\star/M_\odot)$   & Uniform                      & {[}4, 15{]}                 \\
                                        & Metallicity ($Z/Z_\odot$) & Logarithmic                  & {[}0.01, 2.5{]}             \\ \hline
Dust Attenuation                        & Model                     & Calzetti (2000)              & $-$                         \\
                                        & $A_V$ (mag)               & Uniform                      & {[}0, 6{]}                  \\ \hline
Nebular Emission                        & $\log(U)$                 & Logarithmic                  & {[}$-4, \, -2${]}           \\ \hline
Redshift                                & $z$                       & Uniform                      & {[}0, 15{]}                 \\ 
\hline \hline
\end{tabular}

\caption{Model priors and parameters for MINERVA UDS-DR0 SED Modelling with \texttt{BAGPIPES}}
\label{tab:Bagpipes_Priors}
\end{table*}

PSFs for all JWST and HST filters were constructed with PyPHER \citep{Boucaud2016} from 4$''$ stamps of stars in the field, at the native pixel scale and with appropriate regularization parameters. All mosaics were PSF-matched to F444W--which has the broadest PSF--before photometry. Aperture photometry was performed with AperPy \citep{Weaver2024} in 0.20$''$, 0.32$''$, 0.48$''$, 0.70$''$, 1.00$''$, and 1.40$''$ diameter apertures. Kron-correction and a wavelength-dependent correction for Galactic extinction was done, in addition to a $5-10\%$ correction from the F444W curve of growth beyond $1''$.

The SUPER catalog adopts the best of the six apertures through adaptive aperture selection based on isophotal areas \citep[following][]{Labbe2003}; the isophotal area for objects flagged as blends was reduced by 20\% to ensure robust colors. Small objects blended with nearby large sources were flagged and measured without Kron photometry, with fluxes scaled to total via the PSF curve of growth. 

Flags identify spurious detections, stars, diffraction spikes, artifacts, marginal detections, and sources without usable photometry. \citetalias{Cutler2026} combined these flags into a master flag, \texttt{use\_phot}, that identifies sources with reliable photometry. We use only sources with \texttt{use\_phot == 1}. The SUPER catalog already accounts for bad pixels or mosaic edges within the apertures by choosing a smaller aperture in some cases.

MIRI fluxes were derived from the color of each MIRI filter relative to NIRCam F444W, measured in a fixed reference aperture on both the PSF-matched F444W image and the MIRI image.

The SUPER catalog provides identifiers, coordinates, photometric fluxes and uncertainties, exposure times, weights, quality flags, basic morphological quantities such as Kron parameters, background-noise parameters, and available spectroscopic redshifts for all sources identified in the MINERVA-UDS field. We pass the photometric fluxes and uncertainties as input to our BAGPIPES SED modeling, which forms the basis of this work.

\subsection{SPS Modeling with BAGPIPES}
\label{sec:bagpipes}

We modeled the spectral energy distributions (SEDs) of the MINERVA UDS-DR0 sources from their photometry--and spectroscopic redshift, if available--using the Bayesian Analysis of Galaxies for Physical Inference and Parameter EStimation (BAGPIPES; \citealt{Carnall2018}) Python package. Using Bayesian stellar population synthesis (SPS) techniques, BAGPIPES generates model galaxy spectra, computes synthetic photometry for the given filters, compares the model spectra with the input data (any combination of photometry and spectroscopy), and identifies the best-fitting spectrum. The posterior distribution is sampled using the MultiNest nested sampling algorithm \citep{Feroz2008, Feroz2009, Feroz2019}, which efficiently and robustly explores the high-dimensional parameter space while concurrently estimating uncertainties; the physical parameters of the best-fitting spectrum are thus returned with their corresponding posteriors. We report $16^{\mathrm{th}}-84^{\mathrm{th}}$ percentile uncertainties (corresponding to $1\sigma$ on a Gaussian distribution) and use them to plot error bars throughout our analysis. BAGPIPES generates spectra from the far-UV to the microwave regimes and includes parametric prescriptions for star formation histories, nebular continuum emission, line emission, dust emission, and attenuation. 

We ran BAGPIPES on the entire MINERVA UDS-DR0 SUPER catalog after selecting sources with robust photometry (\texttt{use\_phot == 1} and \texttt{USE\_PHOT\_MIRI == TRUE}). We adopt nearly the same priors as \citetalias{Biswas2026}, who studied a similar HST-Dark population in the UNCOVER field. We keep the priors as uniform as possible within physically viable limits to avoid biasing the modeling, while ensuring physically consistent results. The models, model parameters, and priors are as listed in Table~\ref{tab:Bagpipes_Priors}. %

We use the 2016 version of the \citet{Bruzual2003} SPS models, which model the UV--optical region using the Medium-resolution Isaac Newton Telescope Library of Empirical Spectra \citep[MILES;][]{FalconBarroso2011}. We adopt the \citet{Kroupa2002} initial mass function for the simple stellar populations, which has a tunable high-mass slope. We use a flexible delayed-$\tau$ star formation history (SFH) to allow a wide range of stellar ages and formation times; in particular, we differ from \citetalias{Biswas2026} by allowing for a $10\times$ lower minimum age for stellar populations. Nebular continuum and line emission are modeled self-consistently with the stellar continuum using the latest version of the Cloudy photoionization code \citep{Ferland2017}, with a prior on the ionization parameter of $-4 < \log U < -2$. We model dust attenuation with the \citet{Calzetti2000} law, allowing the V-band attenuation $A_V$ to vary over a wide prior range of 0--6. Self-consistent dust and nebular modeling is crucial for this population, which we expect to contain a large fraction of massive, dusty galaxies. Such rest-UV-invisible galaxies could possibly contain unusually large amounts of dust, motivating the high upper limit of the $A_V$ prior; however, strong nebular line emission and the broader dust continuum emission are strongly photometrically degenerate and require careful modeling.

We thus fit photometry in 10 HST filters, 19 JWST/NIRCam filters, and 4 JWST/MIRI filters from PRIMER-UDS, MINERVA-UDS DR0, and other observations of the field, as given in the SUPER catalog (Section~\ref{sec:photometry}; \citetalias{Cutler2026}; \citetalias{Muzzin2025}). To account for residual systematic and calibration uncertainties, we add a 5\% error in quadrature to the SUPER catalog uncertainties, following the prescription of \citet{Boyer2022} and \citet{Wang2024}.

%% file: morphology.tex
\subsection{Non-Parametric Morphological Fitting with \texttt{statmorph}}

\texttt{statmorph} \citep{RodriguezGomez2019} is a Python package for non-parametric morphological analysis of galaxy images. Given the image, segmentation map, weight map, and PSF of a source, it measures Kron parameters, orientation, dimensions, several light radii, Gini--M20 values, Concentration-Asymmetry-Smoothness (CAS) statistics, and other quantities without assuming a profile for the radial distribution of galaxy light. We ran \texttt{statmorph} on the F356W images of the entire MINERVA population of $\sim$180,000 sources with usable photometry (\texttt{use\_phot == 1}), using a custom multi-core pipeline described in Section~\ref{sec:statmorph-pipeline}.

From the \texttt{statmorph} outputs, we use the 80\% light radius, $R_{80}$, which is defined as the radius of the circle containing 80\% of the total integrated flux of a source (as measured within an aperture of radius $1.5 \,R_{\mathrm{Petrosian}}$), centered at the point which minimizes the asymmetry of the source. Given the proportionality between stellar mass and luminosity (across the stellar mass function), $R_{80}$ traces the dense bulk of the galaxy since the stellar mass distribution well-approximates the total baryonic mass distribution. The missing 20\% of light is a sub-dominant contributor to the uncertainty compared to the stellar mass function modeling uncertainties, and the low-density outer regions most affected by cosmological dimming, resulting in erroneous photometry, lie outside $R_{80}$, making $R_{80}$ a reliable tracer of the baryonic distribution.

Although it specializes in non-parametric measurements, \texttt{statmorph} can also perform single-S\'ersic or double-S\'ersic (bulge--disk decomposition) fits to limited accuracy. Reliable parametric modeling that accounts for all the subtleties of such fitting requires dedicated packages such as GALFIT \citep{Peng2002, Peng2010jun_morfit}. We nonetheless record the \texttt{statmorph} single- and double-S\'ersic fits for qualitative morphological studies in future work. In this paper, we use only R$_{80}$ as a proxy for source size.

\subsection{The \texttt{statmorph} Pipeline}
\label{sec:statmorph-pipeline}

Our Python pipeline first loads the F356W image, inverse-variance weight map, WCS, segmentation map, and official MINERVA F356W PSF (\citetalias{Muzzin2025}; \citetalias{Cutler2026}), all given in FITS format. The MINERVA PSFs are constructed by stacking bright stars in the field (Section~\ref{sec:photometry}). The pipeline loads the IDs and R.A./Dec. coordinates of all $\sim$180,000 sources from the MINERVA UDS-DR0 SUPER photometric catalog and locates each source using the WCS. After this global loading step, multiple independent workers (server cores) process the given sources in parallel.

For each source, the pipeline uses the segmentation map to adaptively select the square cutout size with side lengths of 2.5$''$, 4$''$, 6$''$, 8$''$, 10$''$, and 12$''$, increasing the size until there are $\geq$ 5 pixels between the segment edge and the cutout edges. Cutouts are centered on the catalog position of the source, i.e., the centroid of the detected segment. The sigma (standard deviation) image is calculated as the inverse square root of the inverse variance weight map. Segments in the segmentation map are labeled with their catalog IDs, and the pipeline reads the ID of the central segment to confirm the source ID.

The pipeline then builds a quality-control table of flags and parameters describing the cutout and segment, including:

\begin{itemize}
    \item \texttt{cutout\_truncated}: True if the segment extends beyond the largest cutout size (12$''$ per side).
    \item \texttt{segment\_pixels}: The number of pixels labeled with the target (central) catalog ID.
    \item \texttt{cutout\_size}: The final cutout dimensions chosen by adaptive selection.
    \item \texttt{touches\_edge}: True if the smallest cutout could not contain the segment within a 5-pixel margin, implying that a larger size is used.
\end{itemize}

We set \texttt{include\_doublesersic = True} to perform both single- and double-S\'ersic fits, and run \texttt{statmorph} on the cutouts of the image, the sigma map (standard deviation of the noise), the segmentation map, and the PSF. In addition to centroids and diagnostic S\'ersic fits, \texttt{statmorph} computes several light-fraction radii, Kron radii, CAS--Gini--M20 and MID statistics. The pipeline appends all scalar \texttt{statmorph} outputs, the success or failure status of each run, any diagnostic error messages defined in the pipeline, and the runtime of each run to the quality-control dataframe, which is saved as the final result.

%% file: sample_selection.tex
\begin{figure*}[t]
    \centering
    \includegraphics[width=\textwidth]{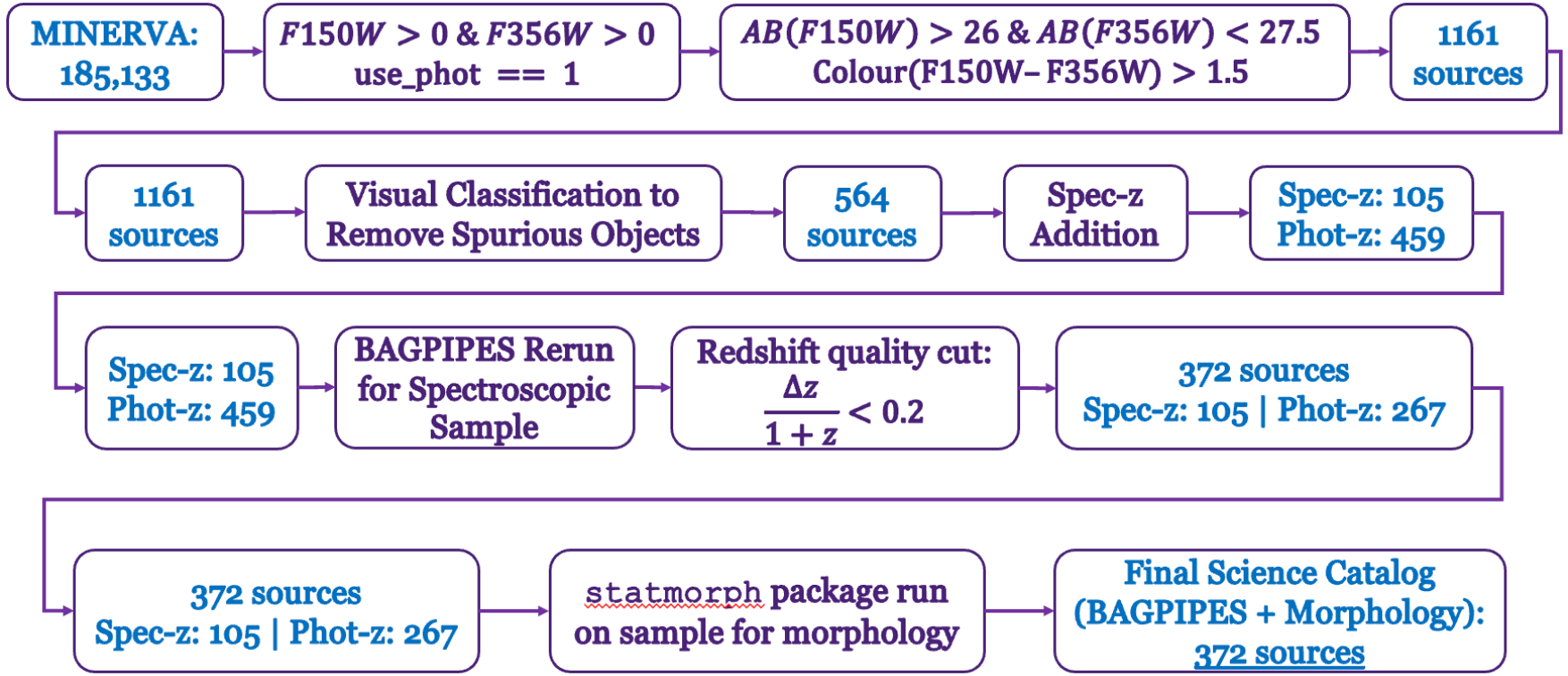}
    \caption{Our sample selection pipeline. The color and magnitude cuts and the 20\% redshift-uncertainty cut follow \citetalias{PerezGonzalez2023}, \citetalias{Barrufet2023}, and \citetalias{Biswas2026}. We perform SED modeling with \texttt{BAGPIPES} \citep{Carnall2018} and non-parametric morphological modeling with \texttt{statmorph} \citep{RodriguezGomez2019}. The final HST-Dark sample contains 372 sources, 105 of which have high-quality (\texttt{grade == 3}) DJA JWST/NIRSpec spectra.}
    \label{fig:sample-selection-flowchart}
\end{figure*}

\subsection{Source Selection}

Our selection method follows prescriptions from recent studies of HST-dark galaxies, primarily \citetalias{Barrufet2023}, \citet[][hereafter \citetalias{Barrufet2026}]{Barrufet2026}, \citet{RodriguezGomez2019}, \citet[][hereafter \citetalias{PerezGonzalez2023}]{PerezGonzalez2023}, and \citetalias{Biswas2026}. The HST-Dark criterion combines a color threshold in the NIRCam F150W and F356W filters with an interval in F356W magnitude to account for both the limited red wavelength coverage and the sensitivity limit of HST, while ensuring robust detections at the (much better but still limited) sensitivity of JWST.

We started from the full MINERVA photometric catalog of 185,133 sources. We applied the \texttt{use\_phot == 1} flag to ensure reliable photometry and removed sources with negative fluxes in F150W or F356W (the two filters used for HST-Dark selection). To keep our sample comparable to previous results, we then adopted the following HST-Dark criteria from the literature:

\begin{itemize}
    \item F150W $\geq$ 26 mag
    \item F356W $\leq$ 27.5 mag
    \item F150W -- F356W $\geq$ 1.5
\end{itemize}

The lower magnitude limit, F150W $> 26$ mag, represents the faintness limit of HST for reliable detections. The upper limit, F356W $< 27.5$ mag, is the NIRCam $\sim 10.44\sigma$ detection limit at the depth of PRIMER, whose broad-band filters we use for selection. The color threshold of 1.50 mag requires sources to be brighter in the JWST filters than at the 1.5\,$\mu$m red cutoff of HST, comparable to the H $- [4.5]$ color used for selection in legacy H-dropout studies using Spitzer/IRAC data (\citealt{Caputi2015, WangT2016}; \citetalias{PerezGonzalez2023}; \citetalias{Barrufet2026}). All these criteria follow \citetalias{PerezGonzalez2023} and \citetalias{Biswas2026}, although \citetalias{Biswas2026} use a deeper upper magnitude limit of F356W $< 28$ mag due to the better sensitivity of UNCOVER. These criteria select 1161 sources. From these 1161 sources, our final sample was constructed through several careful rounds of sample selection, as summarized in the flowchart in Figure \ref{fig:sample-selection-flowchart}.

\begin{figure} 
    \centering
    \includegraphics[width=\columnwidth]{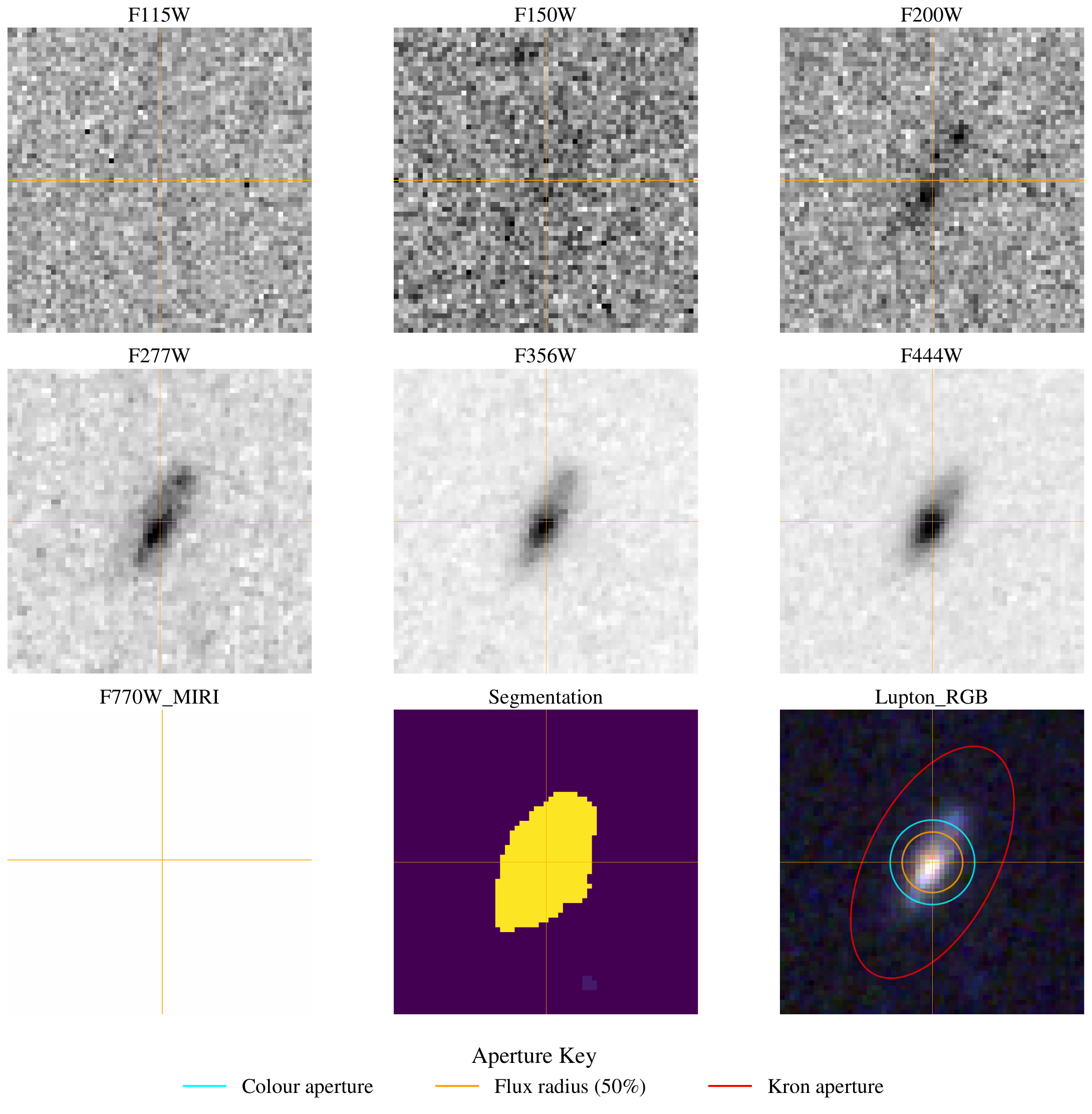}
    \caption{An extended, inclined HST-Dark galaxy (\textit{MinID \# 1003985}). The photometric color apertures are circular, but the Kron correction uses elliptical Kron apertures aligned with the source to minimize the background included in the correction. Kron radii can thus serve as a morphological proxy in uncrowded regions with good SNR. This particular galaxy, however, does not survive the 20\% redshift-uncertainty cut and is hence excluded from the final sample.}
    \label{fig:Mosaic_6_1003985}
\end{figure}

\begin{figure} [tb]
    \centering
    \includegraphics[width=\columnwidth]{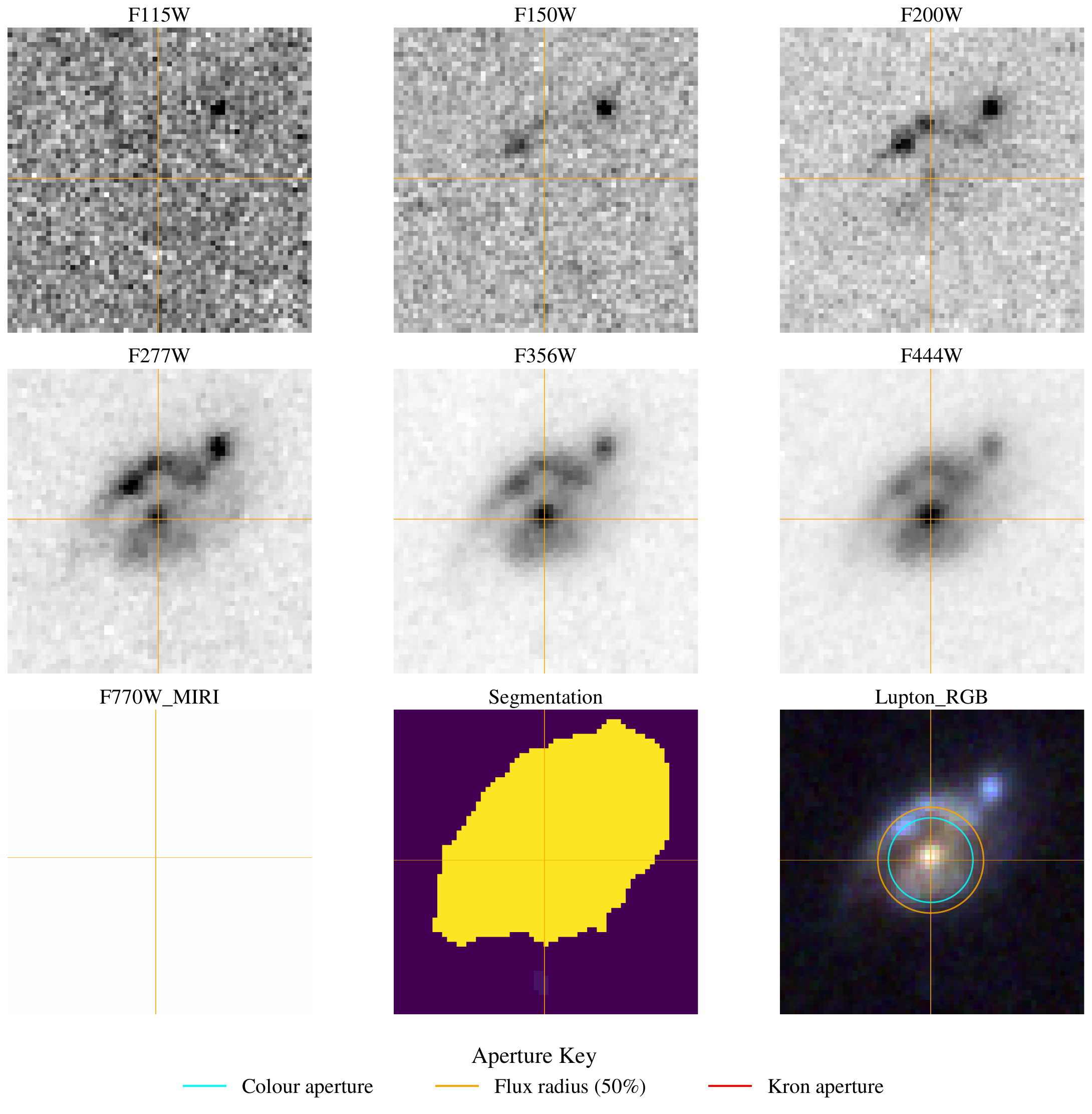}
    \caption{A morphologically interesting HST-Dark galaxy (\textit{MinID \# 1002501)}: an extended system with an apparent spiral-arm-like structure. The redder core is HST-Dark, while bright blue star-forming clumps (likely OB associations) towards the outside are HST-bright. This source also fails the relative redshift cut and is hence excluded from the final sample, akin to \ref{fig:Mosaic_6_1003985}.}
    \label{fig:Mosaic_4_1002501}
\end{figure}

\begin{figure} [tb]
    \centering
    \includegraphics[width=\columnwidth]{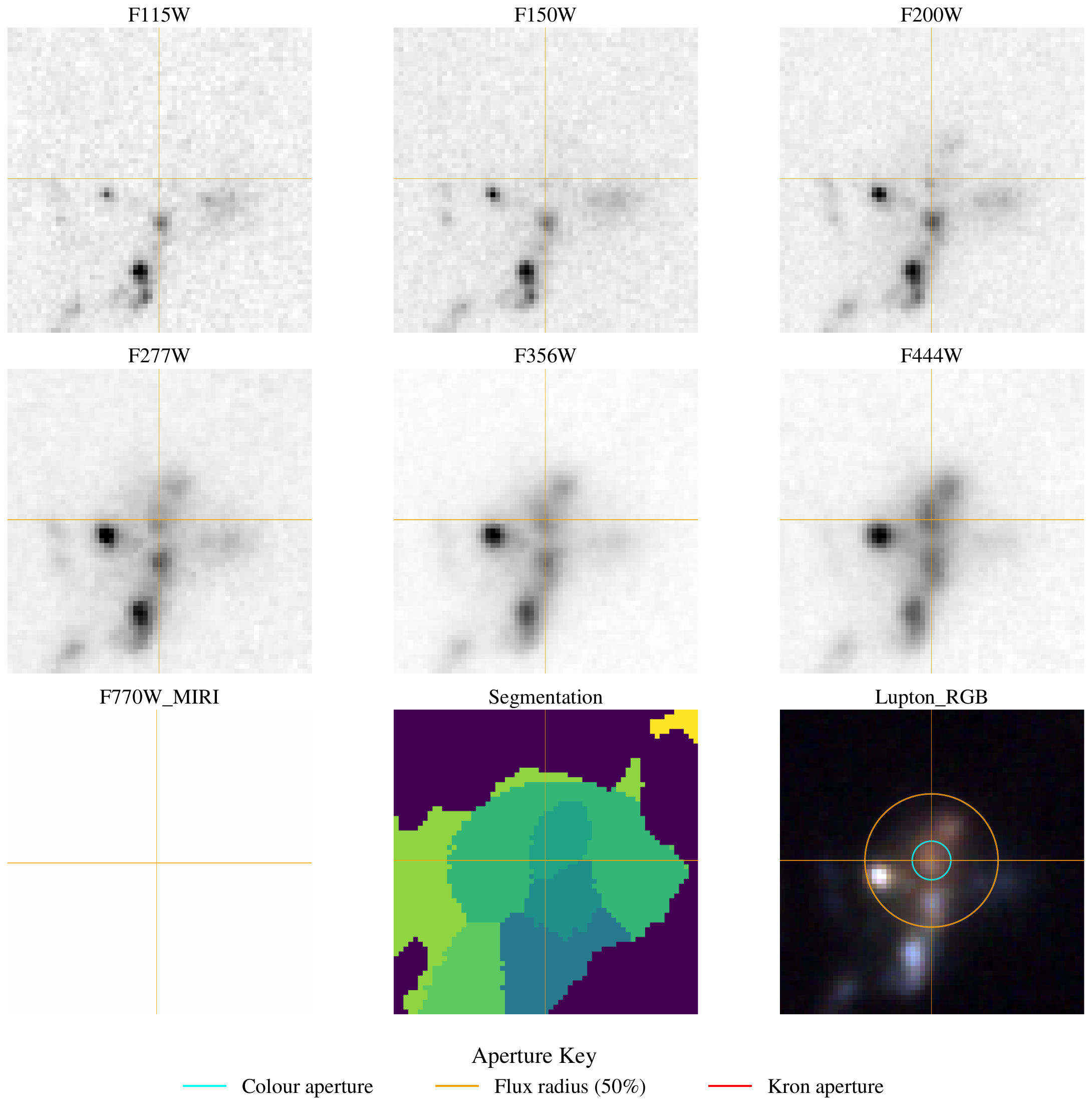}
    \caption{An extended, large, irregular HST-Dark system (\textit{MinID \# 1010916}). Bluer clumps are detected at HST wavelengths, while JWST reveals additional red cores and inter-core gas. The depth and wavelength coverage are insufficient to distinguish between a merging system and a large extended disk affected by cosmological surface-brightness dimming, but all substructures have similar $z$, confirming tidal/morphological interactions. The segmentation map illustrates the aggressive deblending done in the photometry pipeline; only the central clump is measured and selected as HST-Dark.}
    \label{fig:Mosaic_10_1010916}
\end{figure}

\begin{figure}[h]
    \centering
    \includegraphics[width=\columnwidth]{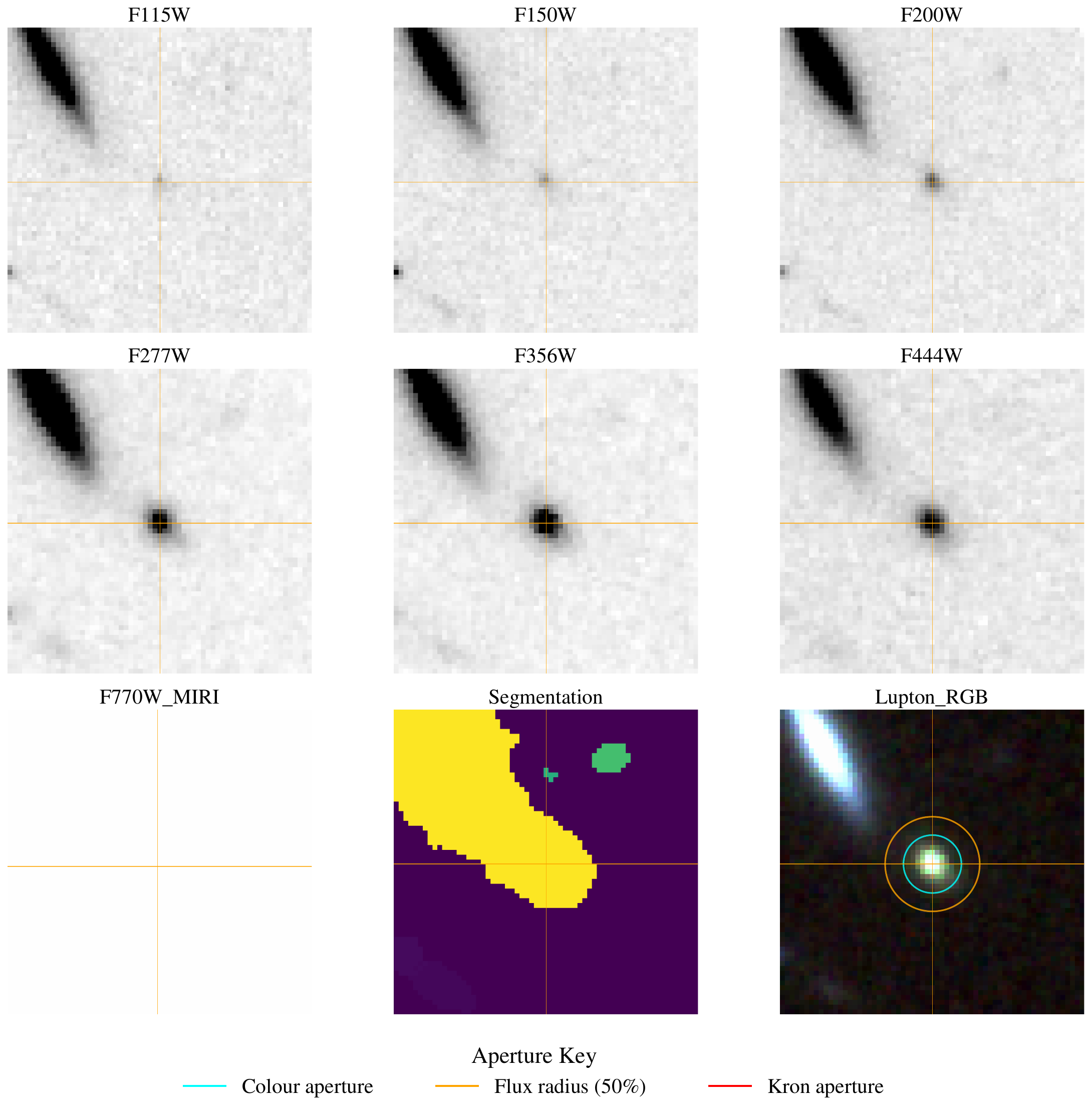}
    \caption{An HST-Dark galaxy (\textit{MinID \# 1004150}) with a bright, prominent neighbor. The segmentation map shows the deblending between the proximal sources. The color aperture and 50\% flux radius show that the photometry is not contaminated, also illustrating the aggressive deblending by the MINERVA team \citepalias{Cutler2026}, even though diffuse emission appears to extend from the neighbor to the central source in F200W and redder filters.}
    \label{fig:Mosaic_7_1004150}
\end{figure}

Despite the extensive efforts of the MINERVA photometry team to remove spurious sources and artifacts, $\sim$50\% of this sample comprised misidentified sources, largely because our targets lie in the faint tail of the source distribution. We therefore visually classified each source using 2.5$''$ $\times$ 2.5$''$ cutouts in 6 NIRCam broad-band filters (F115W, F150W, F200W, F277W, F356W, F444W) and 1 MIRI filter (F770W), as seen in Figures~\ref{fig:Mosaic_2_1018752}, \ref{fig:Mosaic_1_1003714}, \ref{fig:Mosaic_6_1003985}, \ref{fig:Mosaic_4_1002501}, \ref{fig:Mosaic_10_1010916} \& \ref{fig:Mosaic_7_1004150}. MIRI data are available only for some sources, given the MIRI footprint and the partial coverage released in DR0.

We performed several rounds of visual inspection, removing confidently identified artifacts in each round and carrying ambiguous sources forward to the next round, where we used additional information for re-evaluation. In later rounds, we reviewed segmentation maps, Lupton RGB images \citep{Lupton2004} and overplotted Kron apertures to identify real sources, also scrutinizing ambiguous sources in the FITSMap \footnote{Link to FITSMap: \url{https://minerva.colorado.edu/uds/DR0/?ra=34.3560779&dec=-5.1404750&zoom=7}} \citep{Hausen2022} released on the MINERVA website, which allows us to inspect multi-band mosaics with adjustable contrast levels. Because HST-Dark sources are undetected/near-invisible at HST wavelengths and bright in typical JWST bands, they become brighter with increasing wavelength. We used this pattern as a visual criterion when a source's existence was ambiguous in any single cutout: Gaussian noise and hot pixels are uncorrelated across images, so weak but consistent trends across several filters corroborated the presence of a real source. We only retained sources that were visible in at least 2--3 of the long-wavelength filters (F277W, F356W, F444W, stacked for the photometric detection image in \ref{sec:photometry}). In borderline cases, we also rejected sources with very low modeled mass and redshift ($\log(M/M_\odot) \lesssim 7.5$ and $z \ll 1$), which do not correspond to our target sample (Section~\ref{sec:results}). Apart from such rare, extreme cases, where the parameter-based rejection is justified by the clear bimodality between the selected populations (Section~\ref{sec:results}), we avoided using the modeled parameters during selection to prevent subjective biases from entering our sample; the sole aim of the visual classification was to separate real sources from spurious ones.

\begin{figure}[t]
    \centering
    \includegraphics[width=\columnwidth]{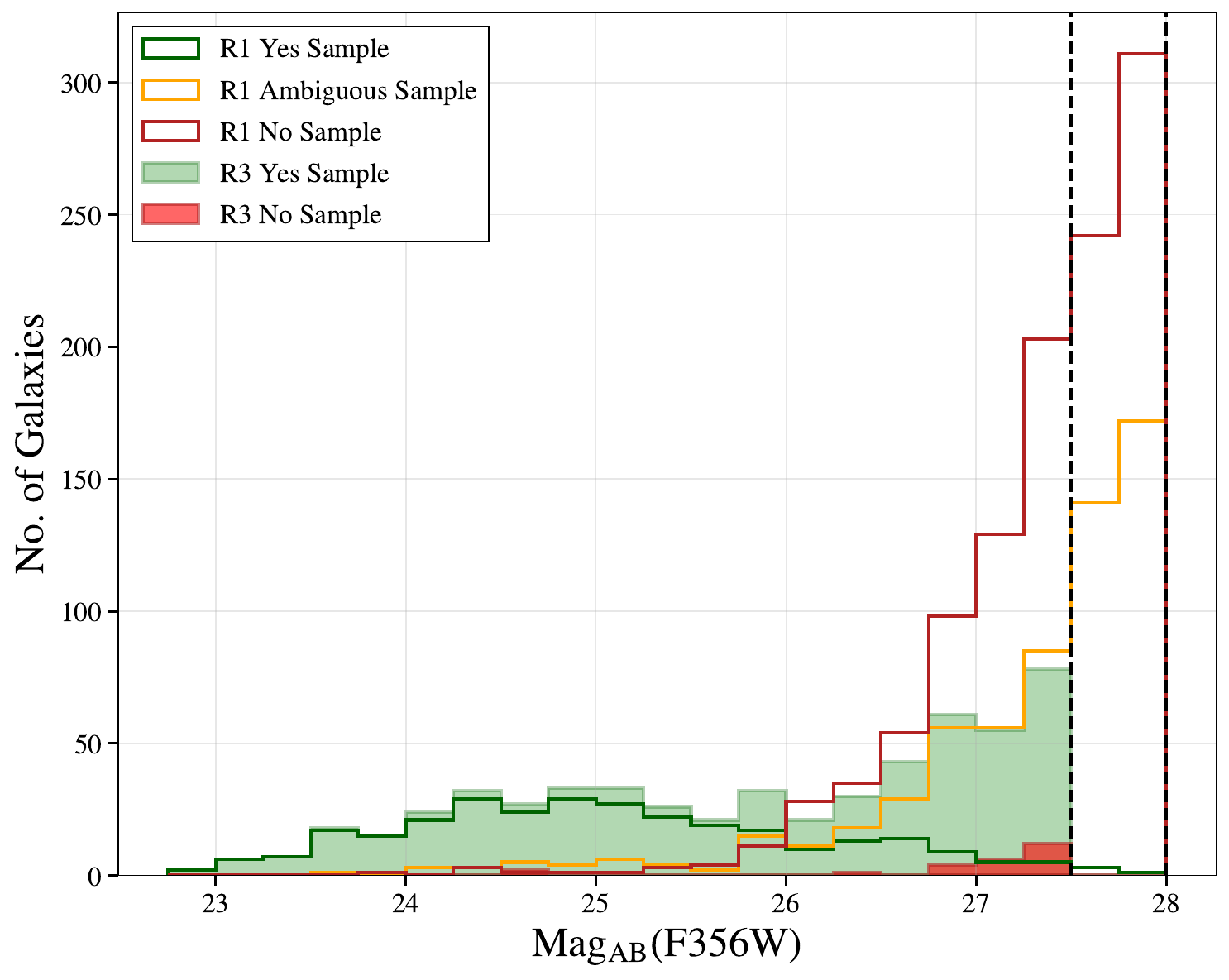}
    \caption{F356W magnitude histograms of all sources with F356W $< 28.0$ satisfying the HST-Dark color criterion; the bin width is 0.25 dex, the 27.5 and 28.0 mag cuts are shown as black dashed lines, and the colors denote the $1^{\mathrm{st}}$ visual classification round results (R1, with F356W $<28.0$) compared with the final round results (R3, with F356W $< 27.5$). Unshaded histograms show the R1 definite-Yes (green), R1 definite-No (red), and R1 Ambiguous (orange) samples. Shaded green and red histograms show the R3-Yes and R3-No samples (R1 definite-No sources were not re-examined, so the shaded histograms sum to R1-Yes + R1-Ambiguous for F356W $<27.5$). The F356W $< 27.5$ cut removed most Ambiguous and definite-No sources from R1 and only 3 definite-Yes galaxies. R3-Yes contains all R1-Yes galaxies with F356W $< 27.5$ (shaded region exceeds the unshaded green histogram). After the cut, most R1-Ambiguous sources (even marginally visible ones) were classified R3-Yes to limit subjectivity; the full R3-Yes sample was then subjected to the $\Delta z/(1+z) < 20\%$ cut for a more objective selection.}
    \label{fig:magnitude-cut}
\end{figure}

We selected the PRIMER broadband filters for visual classification; the PRIMER depths are sufficient to detect sources robustly, while some of the fainter sources are not well-detected in medium bands. We initially used an upper faintness limit of 28 mag in F356W (PRIMER's 5$\sigma$ depth is 28.3 mag, leaving a margin above the detection limit), which selected $\sim$2000 galaxies for visual classification. However, the number of ambiguous or spurious sources was very high, and the inspection became very subjective, so we finally adopted the $\sim$10.4$\sigma$ depth of 27.5 mag. Figure~\ref{fig:magnitude-cut} compares the two limits: in the right panel, unshaded histograms correspond to the 28 mag cut and shaded histograms to the 27.5 mag cut, and R1 and R3 denote the rounds of visual classification. Relative to our initial 28 mag classification, the stricter cut rejected only 3--4 sources from the robust-yes sample (the small tail of the unshaded green histogram beyond 27.5 mag), while reducing both spurious and ambiguous sources by $>50\%$. This confirms that the cut does not remove prominent sources detected well above the survey flux limit (proximity to which also degrades the SED fits). We inspected all remaining ambiguous sources with the checks above to determine whether they were real; regions where the shaded green histogram exceeds the unshaded green histogram show ambiguous objects that were reclassified as real. We flagged a few unusual objects as ``Intriguing'' for later reference.

Many sources appeared very extended or seemed to be merging systems. We flagged these for later photometric inspection if the SUPER catalog aperture extended beyond the apparent visible extent of the source or included nearby sources. Some sources appeared undersampled, with a SUPER catalog \texttt{best\_aperture} much smaller than the object, which were also flagged. However, most of these sources were retained since the aggressive deblending by the MINERVA team separates even close neighbors, and Kron photometry and curve-of-growth corrections account for missing light in apparently undersampled cases. For very compact sources near much larger ones, careful inspection showed that the \texttt{best\_aperture} contained only the target, not the brighter neighbor. The FITSMap was often decisive in such cases, since its multiband color mosaics and adjustable stretch greatly eased the visual classification.

Because visual classification near the detection limit is imperfect and subjective, we biased our selection towards false negatives so that all final sources are reliably real galaxies, at the cost of missing some ambiguous sources. This does not significantly affect global quantities such as the integrated CSFRD, since these low-luminosity galaxies near the detection limit have very low SFRs and masses. Hence, the parameter histograms (Figure~\ref{fig:parameter-histograms}) in this paper describe the prominent population at the available survey depth, with a higher minimum-flux limit (lower magnitude limit) chosen to minimize the Malmquist bias.

After this comprehensive visual examination, we were left with 564 robust, real sources satisfying our selection criteria. The SUPER catalog contains spectroscopic redshifts ($z_{\rm spec}$) from JWST/NIRSpec spectra recorded in the DJA for some of these sources, observed by the programs listed in Section~\ref{sec:data}. We adopted $z_{\rm spec}$ from grade 3 (high-quality) spectra, available for 105 of the 564 sources, and re-ran BAGPIPES with the redshift fixed to these spectroscopic values for these 105 sources. 

\begin{figure}[tb]
    \centering
    \includegraphics[width=\columnwidth]{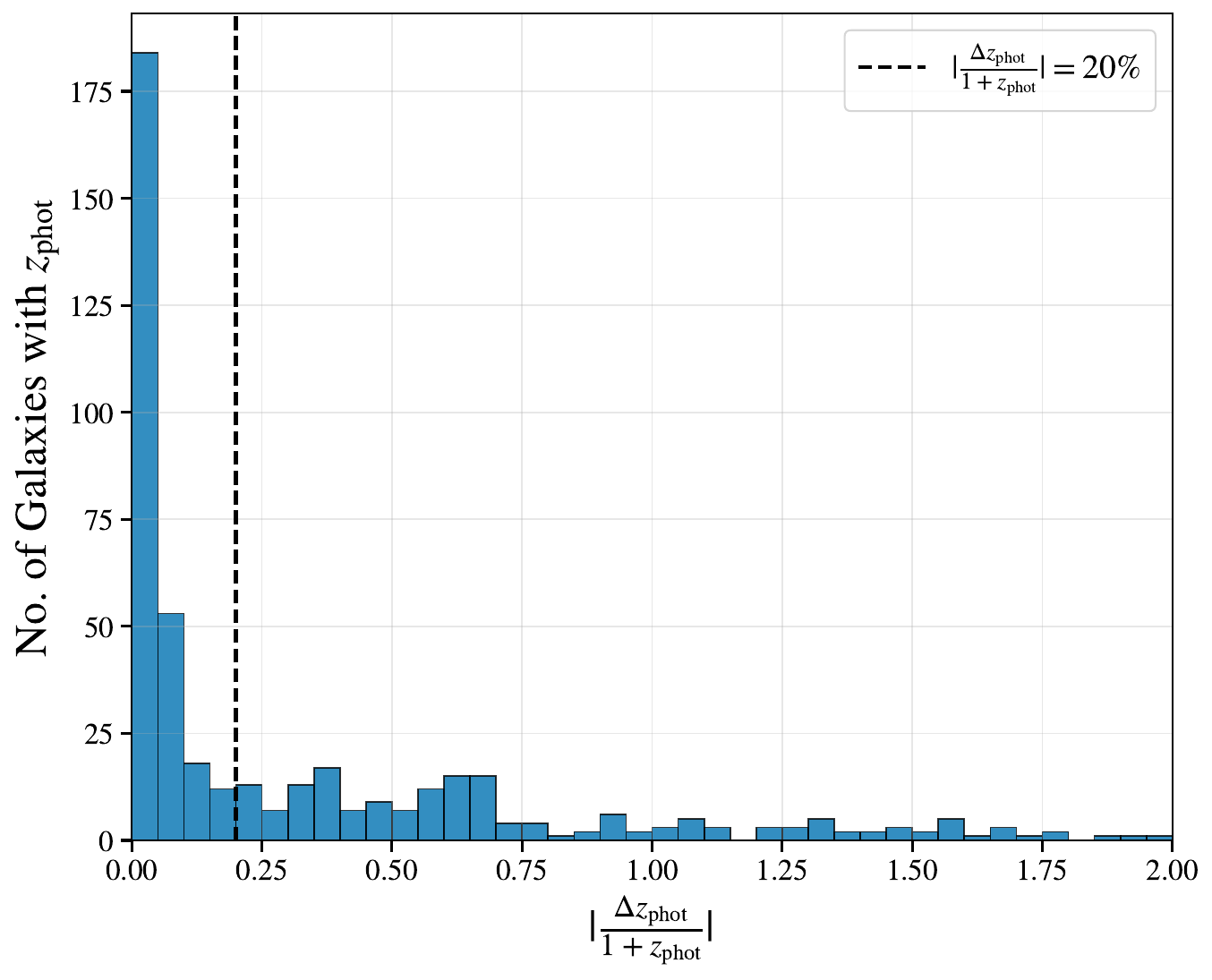}
    \caption{Distribution of (greater of the 2) relative redshift errors for the HST-Dark sample surviving visual classification. The 20\% upper selection threshold is shown as a black dashed line.}
    \label{fig:redshift-relative-error}
\end{figure}

Finally, we applied a quantitative goodness-of-fit criterion to the BAGPIPES redshift posteriors for the entire sample, $\Delta z/(1+z) \leq 0.2$, where $\Delta z$ is the $16^{\rm th}-84^{\rm th} \, \%\rm ile$ redshift uncertainty. Figure~\ref{fig:redshift-relative-error} shows the distribution of relative redshift errors for the 564 sources. Due to asymmetric error bars, the higher percentage uncertainty for each source is considered for the cut and for the histogram plot. This cut removed most of the strongly bimodal redshift posteriors, but it is not completely unbiased: it allows larger uncertainties at higher $z$, as a result of which roughly 20\% of the remaining galaxies show split peaks or weak bimodalities in their redshift posteriors. Since all major (catastrophic) bimodalities have been removed, the remaining uncertain sources are mostly confined to the high-$z$ HST-Dark regime at either redshift solution, and we retain them. Their photometric SED fits are more prone to errors, but such outliers stand out with large error bars in our plots; additionally, most quantitative results are reported as median values to avoid biasing from these extreme outliers. 

Future studies should develop and apply a more robust, unbiased goodness-of-fit test, since this relative cut is disproportionately lenient for high-$z$ objects. We nonetheless adopted it as a quantitative, unbiased test to remove a large fraction of bimodalities and unreliable models that persisted after visual classification, and observed that most of the borderline/marginal/less-robust detections with low SNR failed this test. This produces a reliable, consistent sample with a homogeneous and quantified uncertainty, making it comparable with literature data as required to test for cosmic variance.

After redshift uncertainty cut, our final sample contains 372 HST-Dark galaxies. All 105 galaxies with spectroscopic redshifts survive the cut since the redshift is tightly-constrained during the BAGPIPES re-modeling. Follow-up spectroscopy should therefore be incorporated into such analyses to improve the quality of modeling; imaging studies like ours thus help detect targets for future spectroscopic surveys. The remaining 267 galaxies in the final sample have only photometric data. 

We constructed a science catalog of the BAGPIPES outputs and photometry for these 372 galaxies and analyzed the parameter distributions (Section \ref{sec:results}). We also added the non-parametric morphological measurements of these sources from \texttt{statmorph} to this final science catalog.

\begin{figure}[t]
    \centering
    \includegraphics[width=\columnwidth]{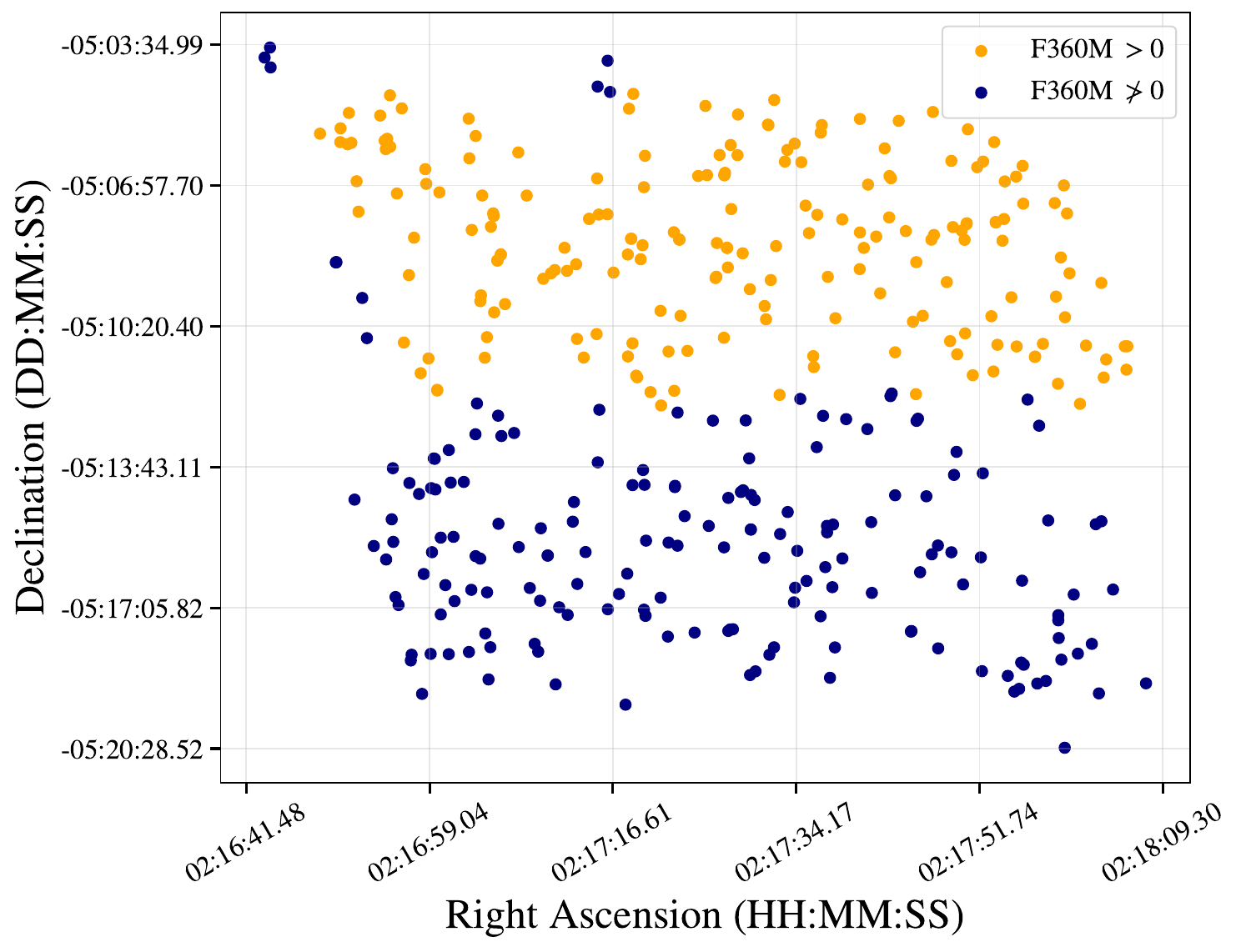}
    \caption{The final HST-Dark sample in R.A./Dec. coordinates. Orange points show sources with MINERVA UDS-DR0 medium-band coverage (F360M $> 0$); navy blue points in the background denote the remaining sources without the DR0 medium band coverage. The upper and lower halves of the field contain similar numbers of galaxies ($\sim 52\%$ with medium bands), indicating no selection effect due to missing medium-band data; we hence retain the full sample.}
    \label{fig:hst-dark-mband-coverage}
\end{figure}

The MINERVA UDS-DR0 release contains medium band imaging for only half of the field but broadband imaging for the whole field (Figure~\ref{fig:hst-dark-mband-coverage}). Our selected sources are nevertheless distributed roughly evenly across both halves: $\sim$52\% of them have medium band data, occupying $\sim$50\% of the survey area. Thus, even after the goodness-of-fit cut, the medium band imaging does not significantly increase the number of detections relative to the broad-band-only region, appearing to have no significant effect on the quality of the fits; thus, we analyze the entire population. The reason for this is unclear and must be investigated in future analyses. It may be symptomatic of the robust depth of the PRIMER broad bands and the very shallow depth of the medium bands, due to which the overall improvements in SED fitting fall short of expectations. Future data releases for other fields will show whether this reflects a systematic limitation of the MINERVA medium-band data or whether PRIMER-UDS, in particular, is deep enough to resolve most SED-fitting uncertainties and degeneracies.

\begin{figure}[t]
    \centering
    \includegraphics[width=\columnwidth]{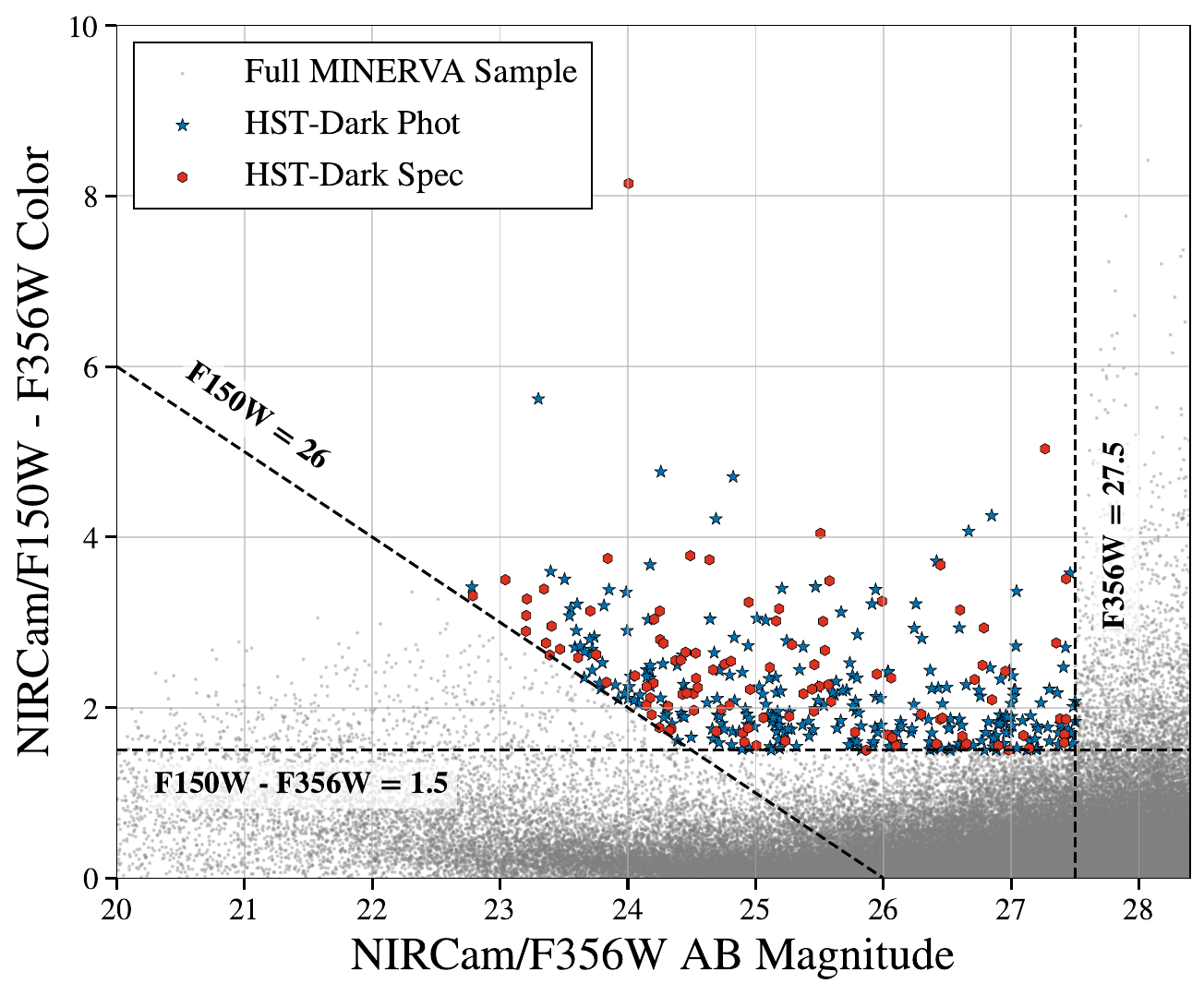}
    \caption{The HST-Dark criterion in F150W$-$F356W color versus F356W magnitude. Dashed black lines show the color threshold (horizontal), the JWST $10.4\sigma$ sensitivity limit (vertical), and the HST sensitivity limit (inclined). Red hexagons indicate galaxies with \texttt{grade == 3} DJA JWST/NIRSpec spectra; the full sample comprises both the red hexagons and the blue stars.}
    \label{fig:source-selection}
\end{figure}

Figure~\ref{fig:source-selection} shows the final sample of 372 galaxies relative to the general MINERVA population and the selection criteria. Galaxies with spectra are shown in red, but the red hexagons and blue stars belong to the same HST-Dark sample and must be considered together when inspecting population-wide properties.

\begin{figure*}[t]
    \centering
    \includegraphics[width=\textwidth]{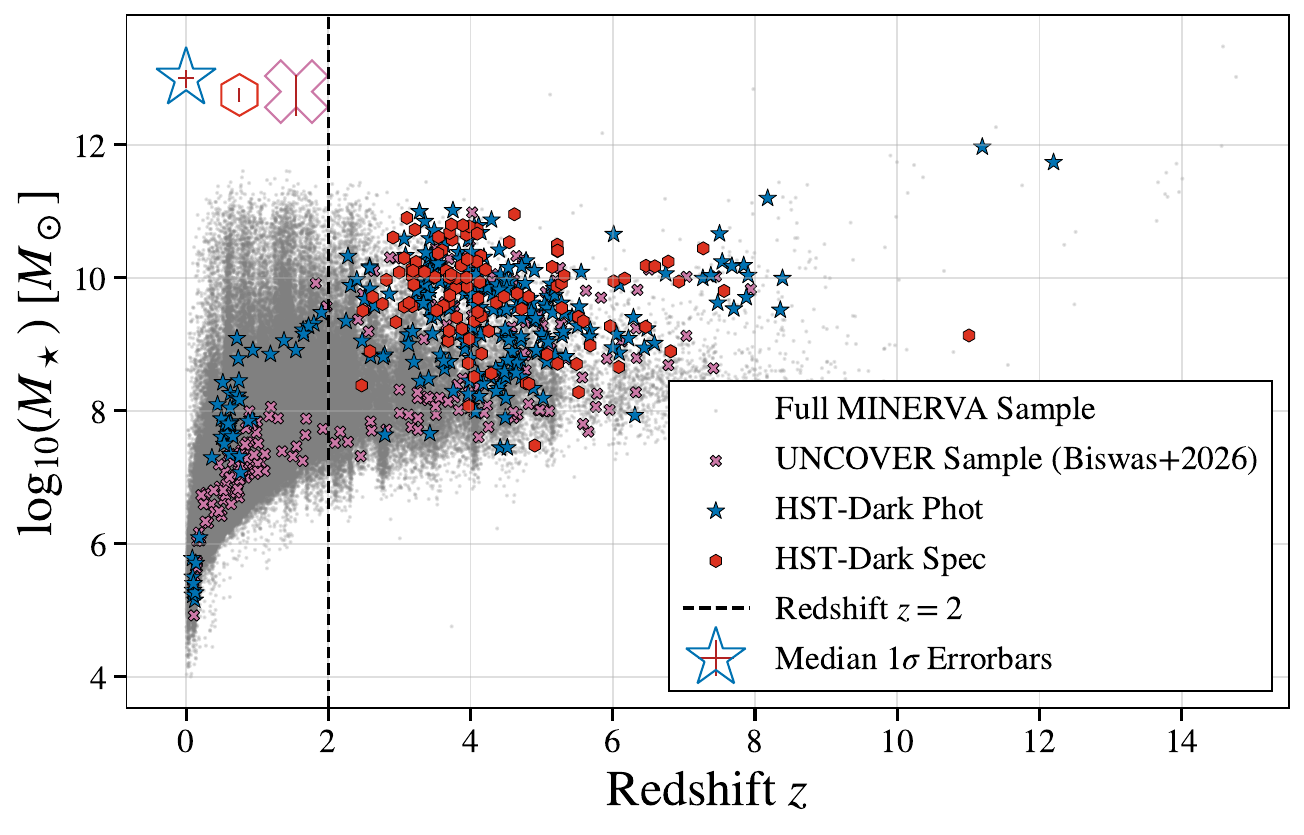}
    \caption{Stellar mass of the HST-Dark population versus redshift, with red hexagons plotted where spectroscopic redshifts are available, and blue stars otherwise. The full MINERVA sample is shown as gray dots and the UNCOVER sources \citepalias{Biswas2026} as pink crosses. Median error bars are drawn in the top corner to avoid overcrowding. Most galaxies lie at $2 < z < 8$. The sample is incomplete at $z < 2$, as shown by the drop in the lower mass limit below $z = 2$, indicating that lower-mass galaxies, although present, are only partially detected. All galaxies with spectra lie at $z > 2$. The systems are predominantly massive, lying above the UNCOVER population and towards the high-mass end of the full MINERVA sample.}
    \label{fig:m-z}
\end{figure*}

\subsection{Blended and Deblended Objects}

Because our sources are faint and distant, their structure is poorly resolved and often entirely below the sensitivity limit. Hence, for compact sources, it is not always possible to determine whether they are independent galaxies, clumps within larger objects, or components of a merging system. Owing to $(1+z)^{-4}$ cosmological surface-brightness dimming, bright OB associations or clumps in a disk may be prominently detected while the disk itself is too faint, thus manifesting as separate neighboring sources. Conversely, an apparent large, merging system may consist of foreground and background galaxies that cannot be segregated in redshift without photometry or spectroscopy of the visible gas halo, tidal streams, etc. Moreover, some parts of a merging or extended system may be HST-Dark while others are not; after aggressive deblending, these may be modeled separately and fit with different redshifts due to the dust in the dark component, adding further uncertainties and degeneracies to the modeling.

Rather than attempting to subjectively diagnose such cases with insufficient information, we treat every selected deblended source as an HST-Dark object or system. In many cases, apparent clumps of a larger system are therefore fit separately as different galaxies, since their identity as individual galaxies cannot be reliably rejected either. Hereafter, ``HST-Dark galaxies'' refers to all such HST-Dark sources. Only rarely were individual clumps of the same visibly larger system selected separately, which supports this choice; overall, $\sim$20\% of our sample consists of objects flagged as presumably merging or extended.

%% file: results.tex
\subsection{Stellar Mass versus Redshift}

 Figure~\ref{fig:m-z} shows the logarithm of stellar mass versus redshift for the full MINERVA sample (gray dots), the UNCOVER HST-Dark sources (\citetalias{Biswas2026}; pink crosses), and our HST-Dark sample (red hexagons where spectroscopic data are available, blue stars otherwise). Both photometric and spectroscopic categories belong to the same sample, and spectroscopic data only yield better-constrained fits; the red hexagons and blue stars should therefore be considered together when analyzing properties and trends. The points represent the median values of the BAGPIPES posteriors. To avoid overcrowding, we show the median $16^{\rm th}-84^{\rm th} \, \%\rm ile$ asymmetric error bars as hollow markers in the top left corner, a convention followed in several subsequent figures. Spectroscopic redshifts have very small uncertainties, which are not shown.

Because MINERVA is a flux-limited survey, it is subject to Malmquist bias: more massive and luminous sources seem to be preferentially detected at higher redshifts, since a large fraction of less luminous sources actually located very close to the detection limit often falls below the observed limit due to statistical fluctuations during measurement. To ensure a homogeneous sample, we impose a completeness limit in redshift. At higher redshifts, the lower boundary of the stellar mass envelope plateaus because lower-mass systems, though present, are too faint to be detected. At lower redshifts, the lower mass limit falls off sharply as less massive systems enter our sample stochastically; including this population would introduce a redshift-dependent selection bias. The fall-off begins at $z \sim 2$ (Figure~\ref{fig:m-z}), which we adopt as the completeness limit. This choice matches the limit used by  \citetalias{Biswas2026}, ensuring a comparable sample, and is not arbitrary: the redshift histogram has a clear minimum at $z = 2$, indicating a bimodal population (Figure~\ref{fig:helm-selection}). No galaxies in the spectroscopic sample lie at $z < 2$. Unless explicitly stated otherwise, we consider only $z > 2$ galaxies hereafter.

This leaves a final sample of 325 HST-Dark galaxies, with a sky surface density of $\sim$1.20 arcmin$^{-2}$, which is significantly denser than any previously studied sample of HST-Dark galaxies except UNCOVER \citepalias{Biswas2026}. 

\begin{figure}[b]
    \centering
    \includegraphics[width=0.9\columnwidth]{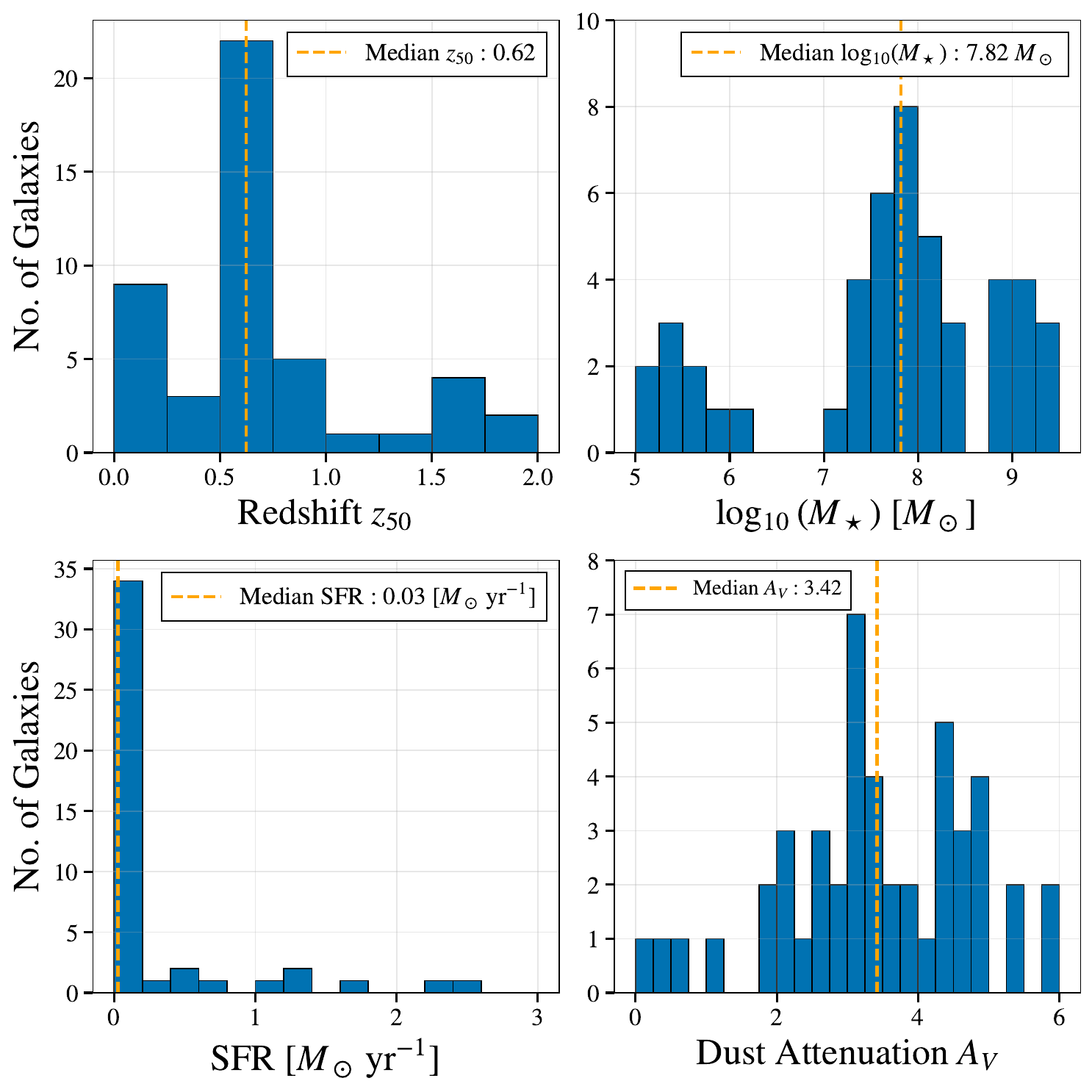}
    \caption{Histograms of redshift, stellar mass, star formation rate, and dust attenuation for the incomplete $z < 2$ sample with 47 galaxies. The discontinuities in the histograms reflect the incompleteness of the sample.}
    \label{fig:low-z}
\end{figure}

\subsubsection*{Low-Redshift ($z<2$) HST-Dark Galaxies}

We first examine the general properties of the $z < 2$ sample.

\citetalias{Biswas2026} find a substantial number of low-mass $z < 2$ galaxies with continuous parameter distributions because gravitational lensing makes UNCOVER imaging effectively much deeper than MINERVA. In contrast, our $z < 2$ sample of 47 galaxies shows discontinuous histograms (Figure~\ref{fig:low-z}), further justifying the exclusion of this incomplete sample from our analysis. This population has a median redshift of $z \sim 0.62$, median stellar mass of $\sim 10^{7.82}\,M_\odot$, median SFR of $\sim 0.03\,M_\odot\,\mathrm{yr}^{-1}$, and median $A_V \sim 3.42$. It therefore consists of proximal, very low-mass, extremely dusty galaxies with minimal star formation. The SFR distribution is almost entirely in the 0--0.2\,$M_\odot\,\mathrm{yr}^{-1}$ bin, and the other three distributions do not show any definite trends.

\citetalias{Bisigello2025} define High-Extinction Low-Mass (HELM) galaxies as sources with

\begin{equation}\label{Eq1:HELM_Criterion_1}
\log_{10}\left(\frac{M_\star}{M_\odot}\right) \leq 8.5
\quad \& \quad
A_V > 1
\end{equation}
\\
\centerline{\textbf{OR}}
\\
\begin{equation}\label{Eq2:HELM_Criterion_2}
\log_{10}\left(\frac{M_\star}{M_\odot}\right) > 8.5
\quad \& \quad
A_V > 1.6\log_{10}\left(\frac{M_\star}{M_\odot}\right) - 12.6
\end{equation}

\begin{figure}[t]
    \centering
    \includegraphics[width=0.85\columnwidth]{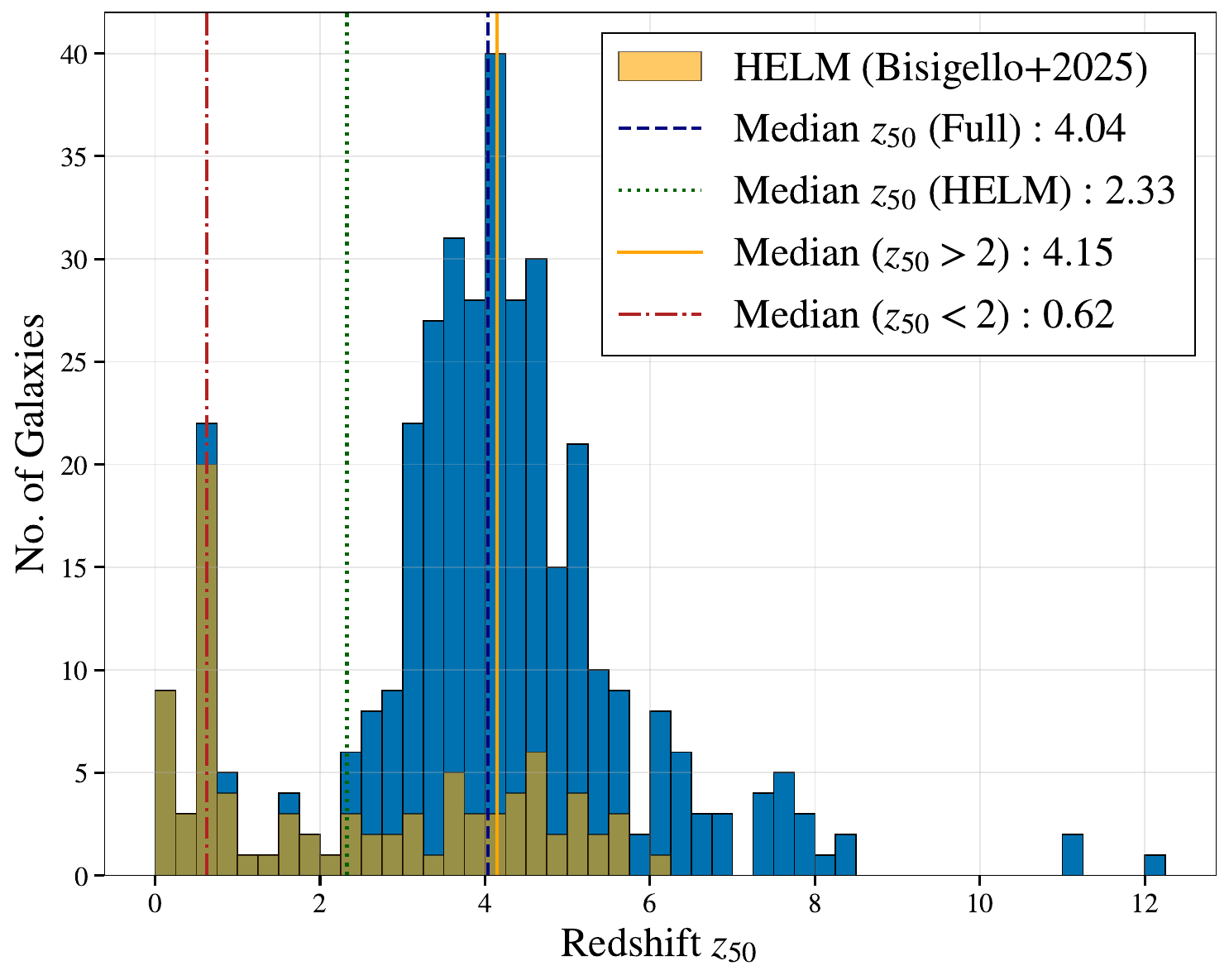}
    \caption{Redshift histogram of the entire population, including $z<2$ galaxies. The minimum at $z \sim2$ shows the bimodality of the population. The HELM criterion of \citetalias{Bisigello2025} is applied to the whole sample and overplotted in orange; most $z<2$ galaxies satisfy it, with a wing extending to high $z$. The median redshifts of the $z < 2$ and $z > 2$ samples, the total sample of 372 sources, and only the HELM selected galaxies is shown using vertical lines.}
    \label{fig:helm-selection}
\end{figure}

We applied this criterion to our entire population and replotted the redshift histogram to show the redshift distribution of HELM galaxies (Figure~\ref{fig:helm-selection}).

\begin{figure*}[t]
    \centering
    \includegraphics[width=0.8\textwidth]{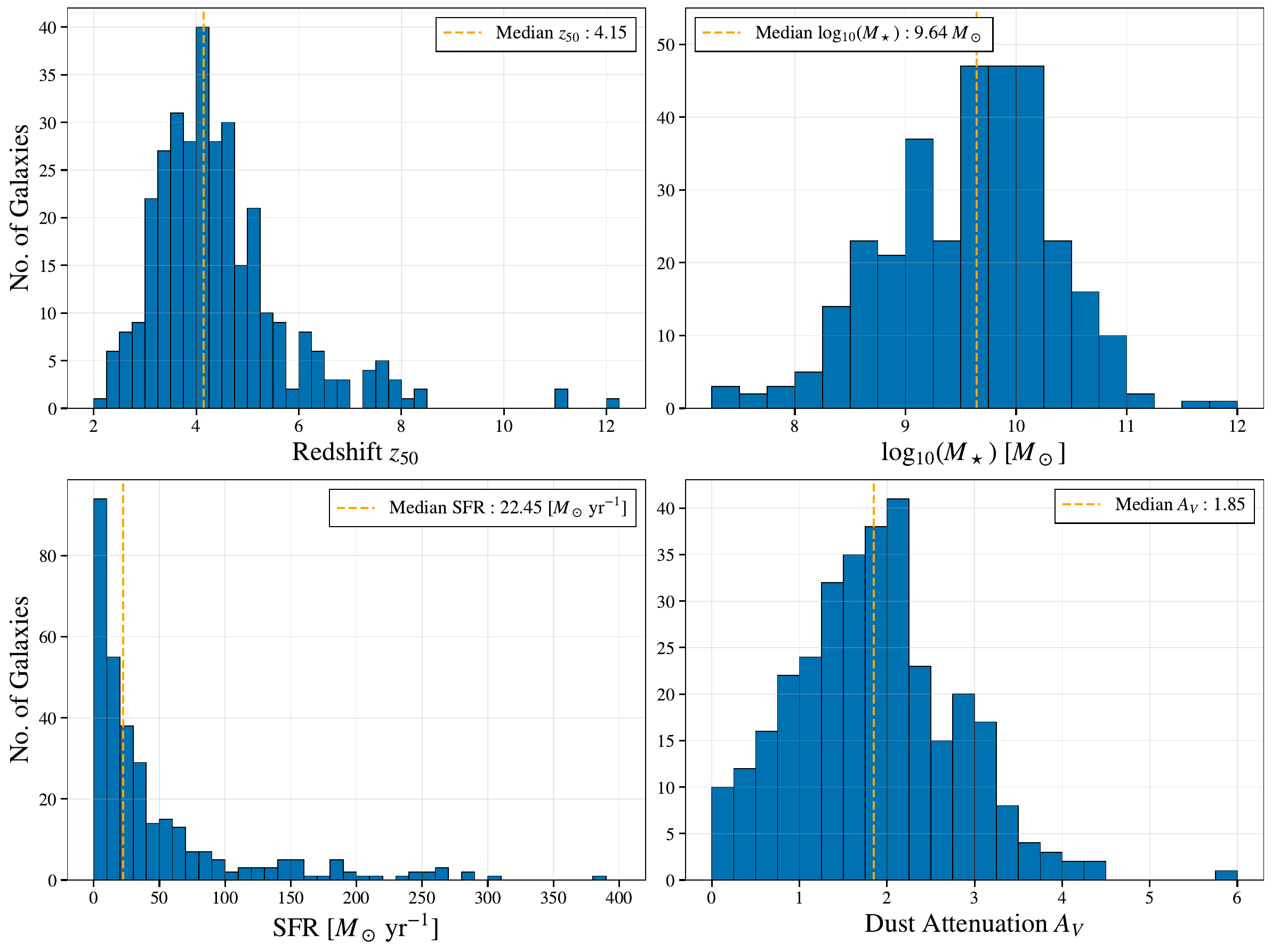}
    \caption{Histograms of redshift, log stellar mass, SFR, and dust attenuation for the final, complete HST-Dark sample of 325 galaxies. The histograms are continuous and clearly unimodal. Orange dashed lines mark the medians, which characterize the sample properties while being robust against extreme outliers, especially for the logarithmically scaled stellar mass. The characteristic population consists of very massive, high-redshift galaxies with large dust attenuation and moderate SFR. The SFR axis is truncated for legibility, excluding a few extreme outliers--we limit the axis limits to 400\,$M_\odot\,\mathrm{yr}^{-1}$, since including these outliers compresses the scale and makes the low-SFR data illegible.}
    \label{fig:parameter-histograms}
\end{figure*}

HELM galaxies extend to $z \sim 6.25$ in small numbers, but all except 4 of the $z < 2$ galaxies satisfy the criterion. Hence, the low-redshift population almost completely demonstrates the typical properties of HELM sources. The HELM classification is not rigid, and \citetalias{Bisigello2025} classify subsets of their sample at 1$\sigma$, 2$\sigma$ and 3$\sigma$ confidence. We do not remove the $z > 2$ HELM sources, which lie within our completeness limit; we apply the HELM classification only to show that, given our HST-Dark criterion and the incompleteness of our survey, this population dominates at low redshift. A natural follow-up is to inspect the available spectra of the $z > 2$ HELM sources.

\subsection{Distributions of Physical Parameters}

We now analyze our final $z > 2$ HST-Dark sample, cleaned of spurious detections and poor fits, in the parameter space of redshift, stellar mass, SFR, and dust attenuation $A_V$ (Figure~\ref{fig:parameter-histograms}). We show logarithmic stellar mass (normalized by solar mass $M_\odot$), $A_V$, redshift and SFR on linear scales, and mark the medians with dotted yellow lines. We use medians to represent typical population properties without being strongly biased by the few extreme outliers.

The median redshift of our sample is 4.15; the redshift distribution has a single dominant mode and is nearly symmetric. The low-redshift tail is truncated by our $z > 2$ cut.

The galaxies are predominantly massive, with a median $\log(M_\star/M_\odot)$ of 9.64. The stellar mass distribution has a clear, broad peak at 9--10.5. The fall-off at the high-mass end is much sharper, as expected physically: in hierarchical galaxy formation, at the characteristic redshift of the population, lower-mass progenitor systems far outnumber higher-mass systems, which have not yet had time to form.

\begin{figure*}[t]
    \centering
    \includegraphics[width=0.8\textwidth]{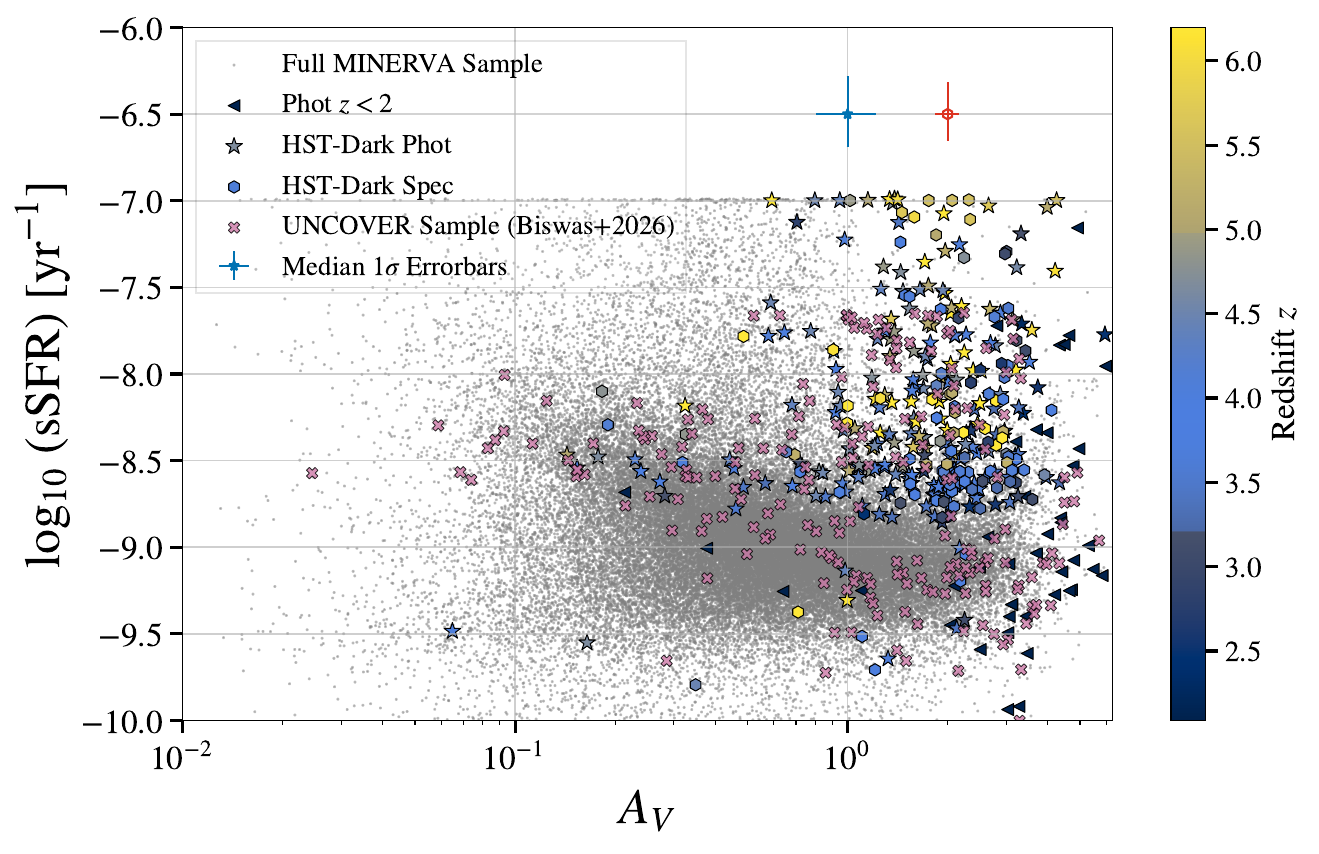}
    \caption{log specific SFR (SFR per unit stellar mass) versus dust attenuation, colored by redshift for the HST-Dark sample. The full MINERVA population is shown as gray dots, HST-Dark galaxies with spectroscopy as hexagons, and the photometric sample as stars; median error bars are shown at the top. The incomplete $z < 2$ sample, shown as left-pointing triangles, lacks completeness and hence does not have characteristic properties; their low-mass, low-SFR (low-sSFR) nature and extremely high $A_V$ resemble typical HELM galaxies. The UNCOVER population of \citetalias{Biswas2026} is added as pink crosses. The fainter, lower-mass UNCOVER population lies at lower $\log \text{sSFR}$ than the MINERVA sample.}
    \label{fig:ssfr-av}
\end{figure*}

The median SFR is a moderate 22.45\,$M_\odot\,\mathrm{yr}^{-1}$: much higher than in local galaxies, but an order of magnitude below the starburst regime. Cosmic star formation activity has been observed to peak at $z > 2$ \citep{Madau2014}, so these systems are expected to have moderate SFRs. The distribution has a clear mode at 0--10\,$M_\odot\,\mathrm{yr}^{-1}$, followed by a steep fall-off. Some extreme outliers reach SFR $>$1000\,$M_\odot\,\mathrm{yr}^{-1}$, indicating extreme starbursts. 

The V-band dust attenuation, $A_V$, has a median of 1.85 and a dominant mode at 2--2.25. The distribution is fairly symmetric, constrained by our $A_V$ prior range of [0,6]. The high-$A_V$ tail falls off slightly more sharply, like the stellar mass distribution, since both arise from hierarchical growth processes with power-law scaling, in which more numerous low-mass systems assemble into higher-mass systems whose numbers fall off beyond a characteristic value.

\subsection{Specific SFR versus Dust Attenuation}
\label{sec:sSFR_AV}

Figure~\ref{fig:ssfr-av} shows the logarithm of the specific SFR (sSFR, i.e., the SFR divided by $M_\star$) against $A_V$, colored by redshift. This plot is central to our analysis, since it shows correlations among all four parameters of the histograms in Figure~\ref{fig:parameter-histograms}.

The HST-Dark sample has a higher dust content and, on average, a higher sSFR than the rest of the MINERVA sample. The higher average sSFR arises from both the high-SFR, high-$z$ galaxies at cosmic noon missed by HST due to redshift and the low-mass, lower-$z$ galaxies whose moderate star formation corresponds to a high sSFR.

Within the $z > 2$ HST-Dark sample, sSFR increases with redshift, reflecting vigorous star formation in less massive systems in the early Universe. There is also a weaker anticorrelation between $A_V$ and redshift, visible most prominently in the densely-populated central region: a brighter, higher-redshift system shifts out of the HST wavelength coverage even without strong attenuation, allowing less dusty systems into the sample. We also expect fewer dusty systems at higher redshift because of the limited time available for dust formation, although studies such as this one are now putting these formation timescales to question. The lower-sSFR population also has a wing extending to low $A_V$. These are the faintest galaxies in our sample, selected either because their relatively low redshift pushes them above the F356W $> 27.5$ limit (as seen below the $z=2$ completeness limit in Figure~\ref{fig:m-z}), or because their sSFR is so low that a combination of very low dust and slightly higher-$z$ is enough to absorb all of their emitted UV light, making them HST-Dark while still bright and red enough for JWST detection.

The $z < 2$ photometric galaxies, shown as left-pointing triangles for comparison, show no clear correlation between dust and sSFR. They are mostly clustered between $3 < A_V < 5$ with similar redshifts, but are distributed across the range of sSFR with a slight overdensity at low sSFR. They therefore do not form a distinct, characteristic population but rather an incomplete sampling of the general galaxy population. Their wide sSFR range reflects a range in low stellar mass values (as seen in Figure~\ref{fig:low-z}, which leads to visible spread since sSFR $\propto M_\star^{-1}$). There are 2 tails to this distribution--the high sSFR tail compensated for by a higher dust content, and another (more typical) low dust tail, where the sSFR is definitely lower to maintain HST-Darkness. These properties are consistent with their identification as HELM sources.

All the data are sharply truncated at $\log_{10}(\mathrm{sSFR}) = -7.0$, an artifact of the BAGPIPES modeling and the definition of sSFR. The stellar populations are modeled with a minimum age of 10 Myr ($10^7$ yr), so in starbursts whose light-weighted age is dominated by stellar populations of this minimum age, sSFR $\propto \rm yr^{-1} \sim 10^{-7}$, i.e., $\log_{10}(\mathrm{sSFR}) \sim -7.0$. A softer lower cutoff appears at $\log_{10}(\mathrm{sSFR}) \sim -8.75$, below which there are very few $z < 2$ HST-Dark galaxies. Sources with very low sSFR have low detected flux, especially due to the absence of many quiescent galaxies in the sample (\ref{sec:UVJ}), thus falling below the detection limit of MINERVA and resulting in this soft cutoff. The presence of a proportional number of quiescent galaxies as reported by \citetalias{PerezGonzalez2023} \& \citet[][hereafter, \citetalias{Barrufet2025}]{Barrufet2025} would have populated the region below this soft cutoff, since a very luminous but massive quiescent galaxy would have low SFR.  

The UNCOVER galaxies have lower sSFR than our sample at every $A_V$, as expected from the better lensing-enhanced depth of that survey. The low-$A_V$, low-sSFR wing is more prominent in UNCOVER, supporting our interpretation that such very-faint sources are a real part of the population near/below the detection limit and are scarce in MINERVA only because of its shallower, incomplete depth, and not because they are intrinsically rare.

\subsection{Star-Forming Main Sequence}

\begin{figure*}[t]
    \centering
    \includegraphics[width=\textwidth]{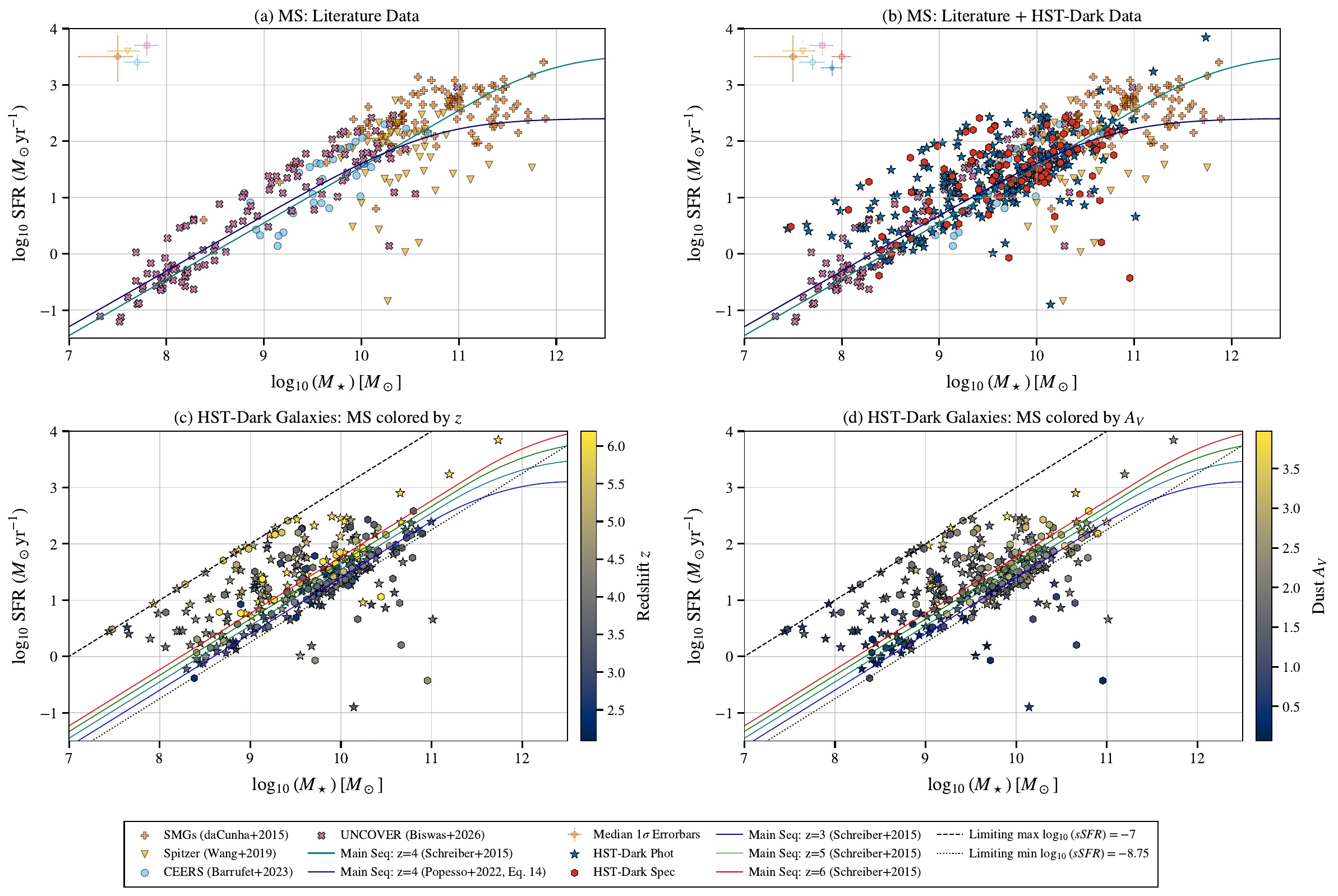}
    \caption{The Star-Forming Main Sequence (SFMS) [$\log_{10}\mathrm{SFR}$ vs $\log_{10}M_\star$], with median error bars plotted in the top left corner. The upper 2 plots [(a) and (b)] show semi-empirical reference plots for the SFMS from \citetalias{Schreiber2015} and \citet{Popesso2023}, drawn for $z=4$. The lower 2 plots [(c) and (d)] show the parametric SFMS plot from \citetalias{Schreiber2015} for a range of redshifts ($3 < z < 6$, with redder colored lines at higher-$z$) to inspect any correlations. The sharp upper cutoff at $\log_{10}\mathrm{sSFR} = -7.0$ (corresponding to BAGPIPES modeling) and the weak lower cutoff at $\log_{10}\mathrm{sSFR} \approx -8.75$ (reflecting the sensitivity limit), as described in Section \ref{sec:sSFR_AV}, are plotted as dashed and dotted black lines. In all plots, the scatter around the SFMS is considerable, but the correlation is nevertheless clear. \textit{(a):} HST-Dark sources from the literature on the SFMS. The semi-empirical main sequence of \citetalias{Schreiber2015} is shown at $z=4$, the approximate median redshift of our sample. The galaxies scatter around the main sequence, but clearly follow the correlation. \textit{(b):} Similar to Figure~(a), with the MINERVA HST-Dark sample added (red hexagons for available spectra, blue stars otherwise). MINERVA fills the gap between the extreme systems observed with ALMA and Spitzer at the high end and the faint limit probed by UNCOVER at the low end, populating the general distribution of CEERS sources with larger numbers. \textit{(c):} MINERVA HST-Dark galaxies on the SFMS colored by redshift, with the spectroscopic sample shown as hexagons and the photometric sample as stars. Higher-$z$ galaxies lie further above the main sequence than lower-$z$ sources, but this spread cannot be explained solely by the upward redshift evolution of the \citetalias{Schreiber2015} SFMS (colored lines), suggesting that updated models are needed. \textit{(d):} The SFMS colored by $A_V$. There is a weak correlation between dust and the position on the SFMS, with higher $A_V$ at the upper end of the SFMS, but no prominent correlation between dust and the scatter about the relation. We observe higher dust content in higher-mass systems.}
    \label{fig:main-sequence}
\end{figure*}

We next examine the distribution of our galaxies on the star-forming main sequence (SFMS), the empirical correlation between log$_{10}$SFR and log$_{10}M_\star$. We draw 2 reference lines for the SFMS--the first is the semi-empirical relation of \citetalias{Schreiber2015} and the second is the relation modeled by Eq. 14, \citet[][hereafter \citetalias{Popesso2023}]{Popesso2023}, both drawn at $z = 4$, approximately the median redshift of our sample (for \citetalias{Popesso2023}'s model, the corresponding age of the Universe measured from the Big Bang was calculated using \texttt{AstroPy}'s Planck18 cosmology module). We first examine the literature populations [Figure~\ref{fig:main-sequence}(a)] before adding our sample [Figure~\ref{fig:main-sequence}(b)].

The ALMA-detected SMGs occupy the extreme upper end of the main sequence, consistent with their nature as extreme starbursts with SFR $\sim$ 1000\,$M_\odot\,\mathrm{yr}^{-1}$ and very high stellar masses, as expected for sources bright enough to be detected in the ALMA continuum from the ground. The Spitzer-selected HST-Dark galaxies occupy a slightly less extreme position. However, the poor angular resolution of Spitzer makes their photometry and SED fits very uncertain: even the median uncertainty in log SFR is too large to fit within the plot, so we omit the y-axis median error bar for this sample. The sensitivity of Spitzer detections is also limited, so these galaxies still do not represent the true median properties of the HST-Dark population.

The JWST CEERS galaxies occupy the central region of the main sequence and are likely a more typical sample of the HST-Dark population. Their uncertainties are smaller, and they follow the main sequence more tightly than the SMGs and Spitzer-selected galaxies. With only 30 sources, however, these trends are indicative rather than statistically significant.

The UNCOVER population forms the low-mass, low-SFR tail of the population, and many of these sources would be undetectable without lensing magnification and the deeper $5\sigma$ sensitivity $\sim 29-30$. A dense grouping of these galaxies extends to very low stellar masses ($M_\star \sim 10^8\,M_\odot$), while the remaining sources populate the main sequence fairly uniformly, with a visible dip in density at $10^{7.5}$--$10^{8.5}\,M_\odot$. Their uncertainties are similar to those of CEERS, with smaller uncertainties for the higher-mass galaxies whose photometry is better constrained due to higher SNR.

The MINERVA HST-Dark population fills the gaps between these samples, as seen in Figure~\ref{fig:main-sequence}(b), spreading fairly uniformly over $10^{8.5}$--$10^{10.5}\,M_\odot$. Most of the spectroscopic sample lies at higher masses, as expected, given the greater difficulty in obtaining spectra for fainter sources. The MINERVA median uncertainties are the smallest of all the samples, comparable only to those of CEERS, owing to the similar depths of the PRIMER and CEERS broad-band imaging. SFR and $M_\star$ are clearly correlated along the main sequence, but the scatter is considerable and cannot be accounted for by the uncertainties alone.

To investigate this scatter, we show only our HST-Dark sample, colored by redshift, together with the semi-empirical main sequence at increasing redshifts [Figure~\ref{fig:main-sequence}(c)]. The main sequence shifts upward in SFR with increasing redshift, as expected from the higher star formation activity in the early Universe. At any fixed stellar mass, our galaxies also follow this trend, with SFR increasing with redshift, as shown by the yellower colors towards the top of the distribution. However, the scatter is still too large to be explained by the evolving main sequence.

\citet{Simmonds2025} use JWST/NIRCam photometry from the JADES program to characterize the SFMS and its scatter between $3<z<9$ based on a sample of galaxies complete down to $\log_{10}(M_\star/M_\odot) \approx 8.1$, which matches the mass distribution of our population. They demonstrate that the scatter in the SFMS is especially susceptible to bursty star formation, with shorter SFR averaging time scales $\sim 10 \rm \, Myr$ corresponding to an intrinsic scatter of $\sigma_{\rm int} \approx 0.4-0.5 \rm \, dex$ that falls to $\sigma_{\rm int} \approx 0.2 \rm \, dex$ at averaging time scales $\approx 100 \, \rm Myr$ across all calculated stellar mass bins. For our modeling with a minimum stellar age (and hence SFR averaging time scale) of $10 \, \rm Myr$, the highest characterized value of the scatter can explain the spread of the dense portion of the sample closest to the SFMS; however, there is a substantial upward scatter of up to $\sim 1-1.2 \rm \, dex$ between the SFMS plot for a given redshift and the corresponding colored data points, especially for $9 < \log_{10}(M_\star/M_\odot) < 10$, which cannot be attributed to bursty star formation alone; thus, we turn to alternative hypotheses.

The main sequence is semi-empirical and was constructed from the data available at the time. The \citetalias{Schreiber2015} relation predates JWST and was calibrated on HST-bright galaxies, systematically missing sources such as ours, which may explain this discrepancy. This further motivates a careful recalibration of these empirical relations using deeper, larger, and more accurately modeled JWST samples. Such discrepancies also show how previously hidden populations such as HST-Dark galaxies can significantly affect our understanding of star formation in galaxies: since the main sequence relates global parameters such as SFR and stellar mass, it informs and constrains smaller-scale star formation models, for instance by constraining the efficiency of gas-to-star conversion. The sharp inclined cutoff at high sSFR (dashed line) is the $\log(\mathrm{sSFR}) = -7.0$ limit described in Section~\ref{sec:sSFR_AV}, and the softer lower cutoff at $\log(\mathrm{sSFR}) \sim -8.75$ (dotted line) has almost no galaxies below it due to selection effects from our survey depth (allowing a small number of sources to stochastically enter the sample).

Coloring the points by $A_V$ instead of redshift [Figure~\ref{fig:main-sequence}(d)] demonstrates that $A_V$ increases with position along the SFMS. Several physical effects may contribute. Higher stellar mass implies more dust production by stellar processes, and a higher SFR (and hence SNe rate) better disperses the produced heavy elements (from the cores of massive stars) throughout the galaxy. Energetic processes such as shocks, jets, and accretion, which drive interactions between stars, gas, and dust, also create more opportunities for complex molecules to form. Finally, there is a selection effect: a galaxy higher on the main sequence is intrinsically brighter and requires correspondingly more dust attenuation to remain HST-Dark.

\subsection{UVJ Diagram}
\label{sec:UVJ}

\begin{figure} [h]
    \centering
    \includegraphics[width=\columnwidth]{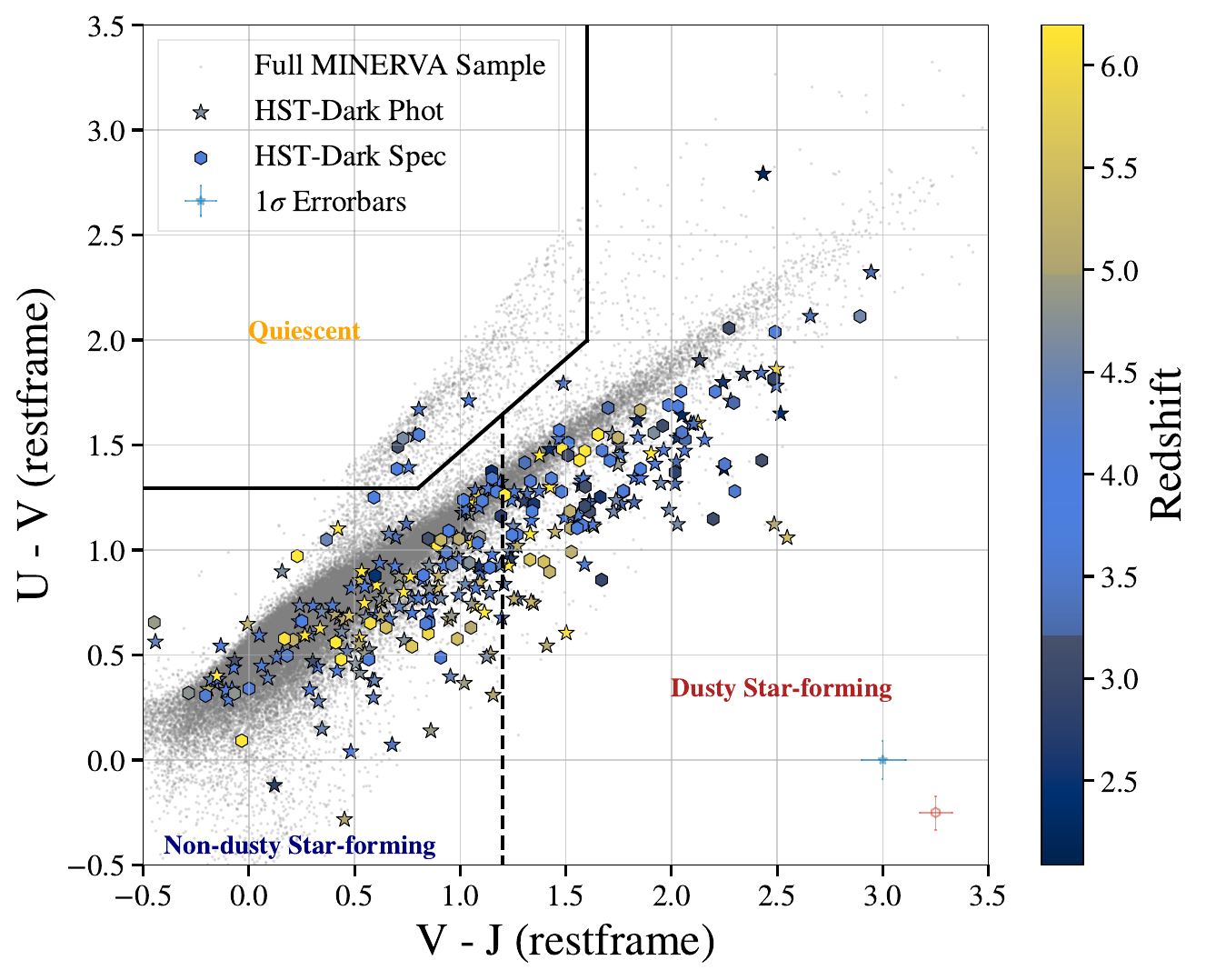}
    \caption{Rest-frame $U-V$ versus $V-J$ diagram separating quiescent from dusty and non-dusty star-forming galaxies. The full MINERVA population is shown as gray dots, and the HST-Dark sources as hexagons (spectroscopic sample) and stars (photometric sample), colored by $z$. The sample contains only 8 quiescent galaxies, all at lower $z$. The higher-$z$ sources are mostly non-dusty star-forming galaxies, and the lower-$z$ sources are dustier.}
    \label{fig:uvj}
\end{figure}

The rest-frame UVJ diagram \citep{Wuyts2007, Williams2009, Patel2012} separates quiescent galaxies from dusty and non-dusty star-forming galaxies. We compute rest-frame $U-V$ and $V-J$ colors from synthetic $U$, $V$ and $J$ photometry performed on the BAGPIPES model spectra and show them in Figure~\ref{fig:uvj}, colored by redshift, with the boundaries between quiescent, non-dusty star-forming and dusty star-forming galaxies taken from \citet{Spitler2014}.

\begin{table*}[t]
\centering
\begin{tabular}{lllll}
\hline
\textbf{Sample}               & & \textbf{Median Redshift $z$} & & \textbf{Median Dust Attenuation $A_V$} \\ \hline
HST-Dark (MINERVA; this work) & & 4.15                         & & 1.85                                   \\ \hline
HST-Dark (CEERS; \citetalias{Barrufet2023})            & & 4.41                         & & 2.06                                   \\ \hline
HST-Dark (UNCOVER; \citetalias{Biswas2026})          & & 4.03                         & & 0.99                                   \\ \hline
SMGs (\citet{daCunha2015})                       & & 2.83                         & & 2.10                                   \\ \hline
\end{tabular}
\caption{Median $z$ and $A_V$ values for sources in Figure~\ref{fig:av-z}}
\label{tab:z_AV_medians}
\end{table*}

Most of our galaxies are star-forming, with only 8 quiescent sources. This is surprising since previous studies (e.g., \citetalias{PerezGonzalez2023, Barrufet2025}) find a larger quiescent fraction ($\sim13-18\%$) among HST-Dark sources at these redshifts. The literature notes that medium-band data is expected to better constrain these sources by sampling emission and spectral features from the older, fainter stellar populations of quiescent galaxies, and breaking degeneracies between quiescent and dusty star-forming sources. It is unclear whether previous studies could not break this degeneracy, leading to a higher quiescent fraction, or whether the MINERVA medium-band depth and quality compromises reliable detections. Future data releases should address this question. Interestingly, the UNCOVER data \citepalias[][]{Biswas2026} reports a quiescent fraction of $\sim 3\%$ compared to our fraction of $\sim2.5\%$, indicating that the true statistically-reliable quiescent fraction might be $\lesssim 3\%$, as opposed to \citetalias{Barrufet2025}'s proposed results from the spectra of only 20 galaxies, subject to significant Poisson errors. A worthwhile follow-up is to inspect the spectra of the MINERVA HST-Dark galaxies classified as quiescent (in Figure~\ref{fig:uvj}) to understand the apparently marked under-representation of this population.

The star-forming galaxies are almost equally divided between the non-dusty and dusty classes. The number of non-dusty star-forming galaxies increases with redshift because HST-darkness results from a combination of faintness and redness: the higher flux due to lower attenuation can be offset by a higher redshift, so that the selection criteria are still met. As expected, the HST-Dark galaxies are redder than the overall MINERVA population, as indicated by their lower $U-V$ and higher $V-J$ colors (the latter sampling the red/infrared regime).

\subsection{Dust Attenuation versus Redshift}

\begin{figure}[t]
    \centering
    \includegraphics[width=\columnwidth]{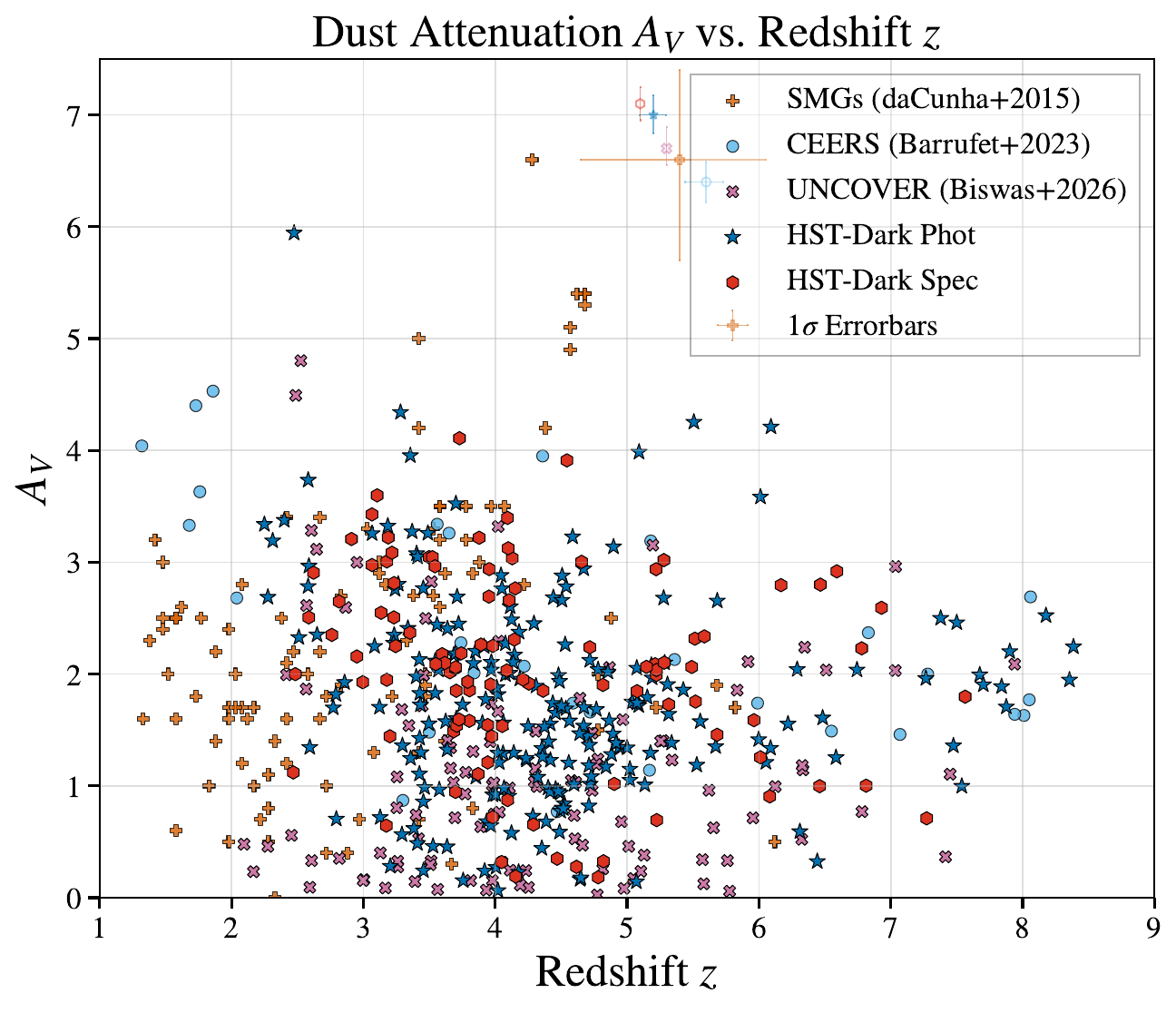}
    \caption{Dust attenuation $A_V$ versus redshift $z$ for HST-Dark galaxies from the literature and from MINERVA. The MINERVA sample is shown as red and blue stars for spectroscopic and photometric redshifts, respectively. Median error bars are shown at the left edge of the legend, behind the data, to avoid confusion with the plotted sources. No prominent relation exists between $A_V$ and $z$, except for the absence of very high-$A_V$ sources at very high $z$, likely due to dust-formation timescales. The scatter shows a substantial presence of dust at early cosmic times.}
    \label{fig:av-z}
\end{figure}

We find no clear correlation between dust attenuation and redshift for our sources (Figure~\ref{fig:av-z}). No galaxies occupy the upper right of the diagram, since very high dust content at very high redshift is unphysical given the timescales required for dust formation. This is also subject to some selection effects, since a source with very high dust and redshift simultaneously may drop out of even the JWST's wavelength coverage and become `NIRCam-Dark', (e.g., \citealp{Perez-Gonzalez2024}). 

The SMGs occupy the low-$z$ region and span a range of $A_V$ extending to high values. This is because ALMA observes the FIR, where dust re-emits the absorbed UV, optical, and NIR starlight, making these sources extremely bright and well-detected in ALMA bands. CEERS extends to higher redshifts, but the sample is too small to show a clear trend. Most UNCOVER galaxies have the lowest $A_V$ and extend to the same redshifts as CEERS, with some at higher $A_V$. The MINERVA sample fills the entire parameter space without apparent correlation or bias, except for a clear density peak near the median values of the distributions, $z \sim 4.15$ and $A_V \sim 1.85$, around which the numbers of sources are highly clustered (Figure~\ref{fig:parameter-histograms}).  

The population medians from Table~\ref{tab:z_AV_medians} are more informative. The SMGs have the lowest median redshift, followed by the deep UNCOVER survey; CEERS has the highest, and MINERVA lies between CEERS and UNCOVER. The median attenuation is highest for the ALMA-detected SMGs (owing to their FIR dust continuum emission, as explained above) and lowest for UNCOVER, which contains low-mass, low-SFR galaxies with little dust. The CEERS sample has a median $A_V$ similar to that of the SMGs, and MINERVA again lies between CEERS and UNCOVER.

\subsection{Cosmic Star Formation Rate Density}

\begin{figure*}[t]
    \centering
    \includegraphics[width=\textwidth]{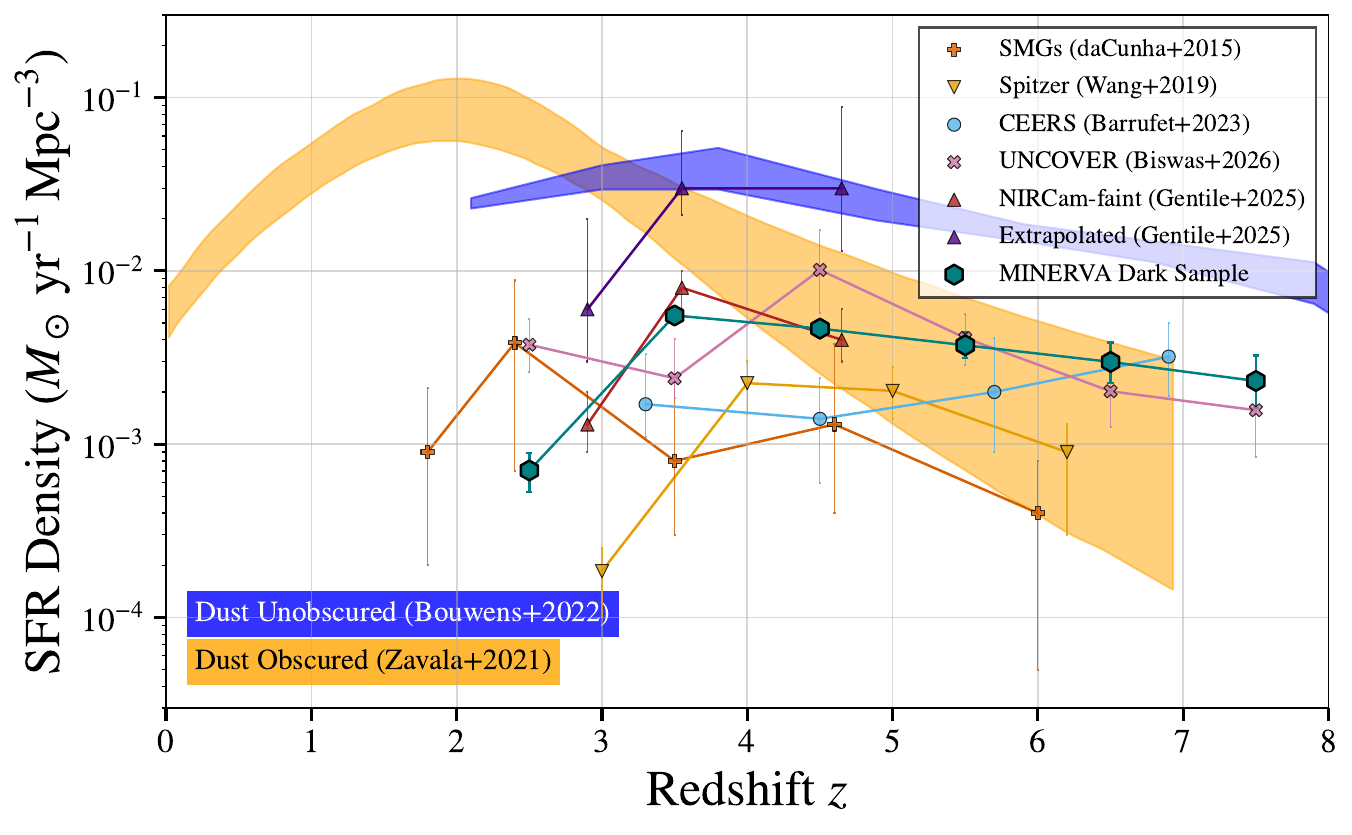}
    \caption{Evolution of the cosmic star formation rate density (CSFRD) with redshift $z$. The blue shaded curve shows the evolution of dust-unobscured CSFRD as presented by \citet{Bouwens2022}, who used recent data to update the trend first demonstrated by \citet{Madau2014}. The orange shaded curve shows the expected evolution of the dust-obscured CSFRD from \citet[][hereafter \citetalias{Zavala2021}]{Zavala2021}, based on scaling factors and semi-empirical calibrations applied to recent data (e.g., IR luminosity functions). We bin the galaxies in $z$ and divide the total SFR of all sources in each bin by the cosmological comoving volume of the bin, plotting the result at the median $z$ of the bin. Error bars depict the quadrature sum of the Poisson sampling errors and the quadrature-summed, volume-normalized SFR uncertainties from the Bayesian posteriors. The MINERVA points reach the upper limit of the dust-obscured model at high $z$, with upper error bars appearing to exceed it at the highest redshifts, contributing $20-30\%$ of the unobscured CSFRD. Our sample is complete only above a certain $M_*$ limit; hence, these results are a lower limit on the integrated dust-obscured CSFRD, motivating deeper observations.}
    \label{fig:csfrd}
\end{figure*}

We next examine the impact of these previously ``missing'' galaxies on global cosmic processes by measuring their contribution to the cosmic star formation rate density (CSFRD) as a function of redshift, which constrains the timescales and mechanisms of mass build-up, chemical enrichment, reionization and structure formation in the Universe.

Figure~\ref{fig:csfrd} shows two shaded semi-empirical curves. The blue curve is the observable dust-unobscured, HST-bright/UV-bright CSFRD as a function of redshift from \citet{Bouwens2022}. This is an updated version of the \citet{Madau2014} relation, constructed from the latest UV luminosity functions after converting UV luminosity to SFR and binning the sources in comoving volume as a function of redshift. The orange curve is the semi-empirical model of \citet[][hereafter, \citetalias{Zavala2021}]{Zavala2021} for the dust-obscured CSFRD (again binned in redshift and normalized by comoving volume), based on ALMA FIR observations of dusty star-forming galaxies, which constrain the evolution of the IR luminosity function and dusty star formation up to $z \sim 7$. The uncertainty of this model grows with redshift because of the scarcity of data, which leads to poor constraints at these redshifts.

The SMGs have extremely high SFRs, but an extremely low sky density ($\sim 0.01$ arcmin$^{-2}$), so their contribution to the CSFRD is very low, with only a handful of sources in each volume bin. Their contribution also declines beyond $z \sim 2$ since they are predominantly low-$z$ sources. The $z = 6$ point touches the lower limit of the \citetalias{Zavala2021} model, but its uncertainties are poorly constrained.

The Spitzer-selected sample lies at higher redshifts and contributes more to the CSFRD, lying well within the dust-obscured model at $z = 5$ and $z = 6$. This is because its $\sim100 \times$ higher sky density compensates for its lower SFR values relative to the SMGs. Its uncertainties are again poorly constrained, owing to the sensitivity and resolution limits of the Spitzer data.

The CEERS sample is the first to challenge the dust-obscured models: with much better-constrained parameters, it lies well within the model at $z = 5.5$ but reaches its upper limit at $z = 7$. \citetalias{Barrufet2023} note that their small sample places only 4--8 galaxies in each bin, so more data are needed to confirm this trend, but it sets lower limits on the CSFRD and motivates further investigation.

UNCOVER lies above the CEERS sample along the CSFRD axis in all redshift bins except the 2 highest ones, with a prominent peak between $4 < z < 5$ where the data reaches $\sim 50\%$ of (and the upper error bar reaches $\sim 60-70\%$ of) the lower limits of the dust-unobscured plot. This illustrates that a significant portion of the CSFRD is missed by dust-unobscured studies, especially at intermediate redshifts; at the highest redshifts, the fall-off may be due to lesser galaxies in the sample due to cosmic variance (since the survey area is still relatively small).

\citet{Gentile2025} use JWST/NIRCam to follow-up on radio-selected NIR-faint galaxies, and model their contribution to the integrated CSFRD across 3 redshift bins. The observed data lies well below/within the orange curve from \citetalias{Zavala2021}; however, they use empirical galactic Schechter functions to extrapolate the properties of their sample beyond their survey limit and calculate complete, global estimates of the parameters. Notably, the extrapolations suggest that the contributions of these galaxies to the total CSFRD could equal that of UV-bright, dust unobscured galaxies at $z\sim4$. Hence, tracing this population and constraining these diagnostic extrapolations is imperative to achieve a complete understanding of early galaxy evolution.

We binned our sources in redshift and computed the comoving volume of each bin in our adopted Planck18 cosmology. The survey area used as input to calculate comoving volume was computed by counting all pixels with non-negative and non-NaN values in the F356W mosaic, and multiplying this number with the area per pixel ($0.04 ''\times 0.04''= \frac{1.6\times 10^{-3}}{3600} \rm arcmin^2$), totaling $\sim 270 \, \mathrm{arcmin^2}$. We summed the SFRs of all sources in each bin, divided the sum by the comoving volume of the bin to obtain the CSFRD, and plotted it against the median redshift of the bin on a logarithmic CSFRD axis. The CSFRD uncertainties are the quadrature sum of the Poisson sampling uncertainty (from the number of sources in each bin) and the quadrature-summed, volume-normalized SFR uncertainties from the Bayesian posteriors.


\begin{figure*}[t]
    \centering
    \includegraphics[width=\textwidth]{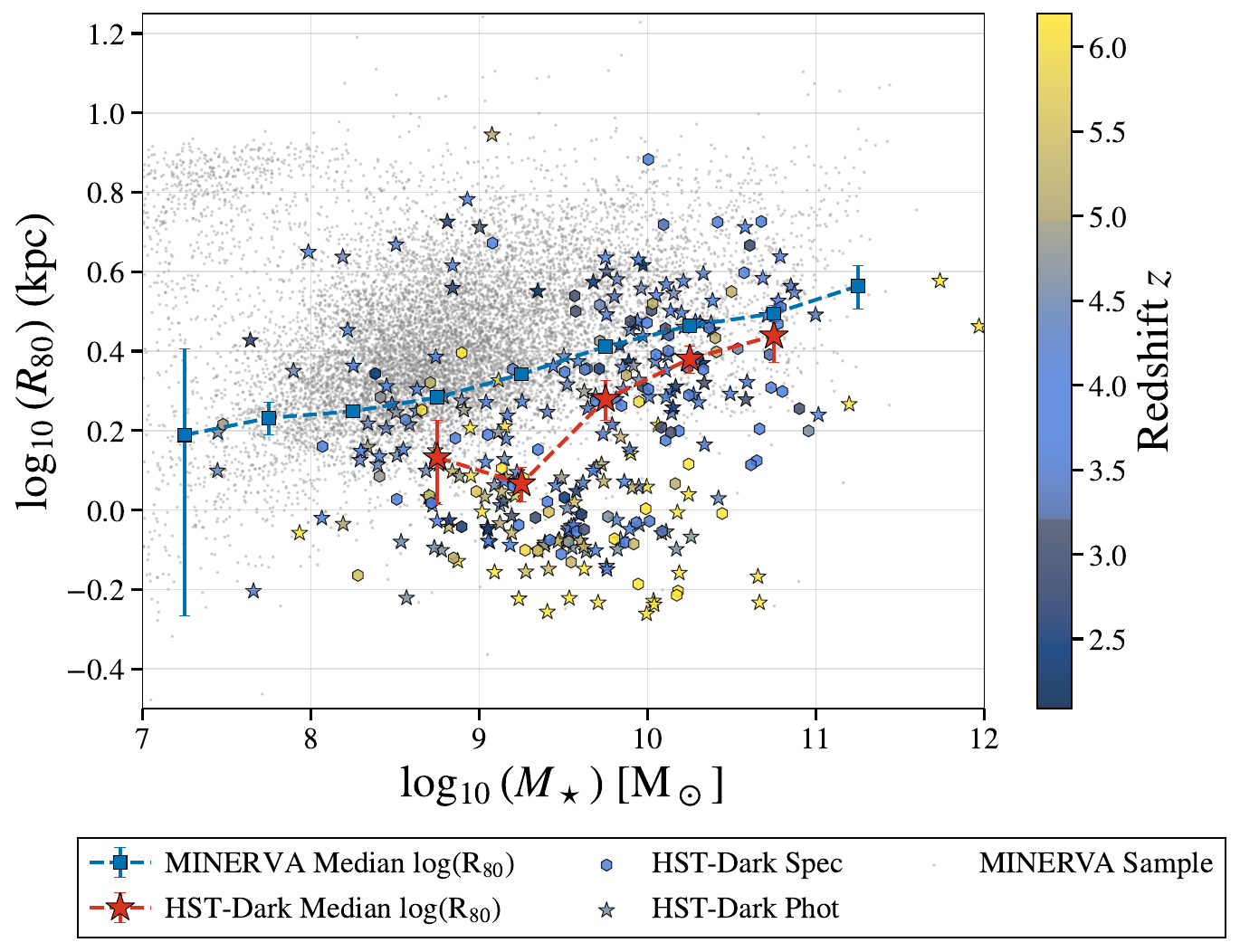}
    \caption{80\% light radius versus stellar mass on a log--log scale. The full MINERVA population is shown as gray dots, and the HST-Dark sample as stars (photometric sample) and hexagons (spectroscopic sample), colored by $z$. For both, the full MINERVA population and the HST-Dark population, we consider galaxies within $3 < z < 5$ (to ensure comparable evolutionary stages), bin them in $\log M_*$ and plot the median $R_{80}$ against the median $M_*$ of each bin, with error bars from 2000 bootstrap resampling realizations. The HST-Dark sources have consistently smaller $R_{80}$ than the broader population, indicating higher baryonic densities. $M_*$ and $R_{80}$ are slightly positively correlated with decreasing $z$, possibly indicating evolutionary links between the higher-$z$ and lower-$z$ HST-Dark sources.}
    \label{fig:r80-mass}
\end{figure*}

The MINERVA population calls into question previous dust-obscured CSFRD models. For $2<z<3$, the SMGs and UNCOVER contribute more due to the extremely high SFR of SMGs and superior depth of UNCOVER, allowing for dominant and numerous detections respectively. At intermediate redshifts ($3<z<6$), only UNCOVER and \citet{Gentile2025}'s samples lie above MINERVA, again due to the superior depths of UNCOVER and the greater sensitivity and spectral coverage of radio-detected NIRCam faint sources, leading to a larger number of samples beyond MINERVA's limits. At the highest redshifts ($6<z<8$), MINERVA has the highest contribution due to the large number of bright sources observed within the large survey area; while the CEERS data has one point above the MINERVA trend, there are only a handful of CEERS galaxies in that bin leading to very significant error bars. MINERVA consistently has the best constrained error bars at all redshifts.

At $z = 4.5$ and above, our points follow the model curve, but the $z = 6.5$ point lies near its upper limit, with the upper error bar reaching the upper limit. The $z = 7.5$ point lies beyond the redshift range of the \citetalias{Zavala2021} model, but would lie at or above its upper limit if the curve were extrapolated smoothly. This is notable because our data are far from complete (due to the survey flux limit, as prominently seen in the lower stellar-mass limit in Figure~\ref{fig:m-z}). Our data are therefore insufficient to make quantitative corrections to the CSFRD model, but they set a strong lower limit on the CSFRD, and a deeper survey would likely raise these points, as demonstrated by the extrapolations performed by \citet{Gentile2025} to simulate completeness and corroborated by the rising contribution of UNCOVER when low-mass sources are sampled. Since low-mass galaxies are far more numerous across redshift and cosmic star formation peaked at higher redshifts, a large fraction of the CSFRD is likely still missing from our census, emphasizing the need to eventually recalibrate these empirical cosmological relations with deeper and more robust data. Most importantly, the CSFRD in the highest-redshift bins approaches $\sim$30\% of the dust-unobscured value, suggesting that deeper future data may motivate a recalibration of the Madau--Dickinson relation. As noted above, this would affect current models of mass build-up, chemical enrichment, structure formation and reionization as a function of lookback time.

\subsection{R$_{80}$ versus Stellar Mass}

Finally, we examine the relation between the physical size of our galaxies and their stellar mass, using the 80\% light radius from \texttt{statmorph}. We convert R$_{80}$ from pixels to arcseconds using the MINERVA mosaic scale of 0.04$''$ per pixel, and then to kiloparsecs using the angular diameter distance at each source's redshift value as per our adopted Planck18 cosmology.

Figure~\ref{fig:r80-mass} shows log R$_{80}$ (kpc) versus $\log(M_\star/M_\odot)$ for the whole MINERVA sample (gray dots) and the HST-Dark population (stars for the photometric sample, hexagons for the spectroscopic sample), with the HST-Dark points colored by redshift. The stellar mass uncertainties are as in the previous figures and are omitted to avoid overcrowding; R$_{80}$ is computed directly from the pixel data and therefore has no Bayesian (modeled) uncertainty.

On average, the HST-Dark galaxies have smaller R$_{80}$ than the rest of the MINERVA sample. R$_{80}$ is also clearly anticorrelated with redshift, with the highest-redshift galaxies having the smallest R$_{80}$ at any given stellar mass. There is a dip in the median $R_{80}$ for the $9< \log_{10}(M_\star/M_\odot) < 9.5$ bin, compared to the overall increasing trend--the reason for this dip is not apparent to us. Hierarchical models of galaxy formation predict that galaxies grow in mass and size with cosmic time, so galaxies should move from the lower left towards the upper right of the diagram as redshift decreases. The bulk of the high-$z$ galaxies lies in the lower half of the plot while the low-$z$ galaxies populate the upper half, demonstrating this diagonal offset in line with these hierarchical models, although the scatter is considerable in the low-$z$, low-mass regime.

To quantify the size--mass relation, we select galaxies at $3 < z < 5$, which are at comparable evolutionary stages, and bin them in $\log M_\star$ with a bin width of 0.5 dex. This is done for both, the entire MINERVA population and just the HST-Dark sample. We compute the median R$_{80}$ in each bin, estimate the sampling uncertainties with 2000 bootstrap realizations, and plot each point at the median $\log M_\star$ of the bin. For the HST-Dark sample, we show bins from 8.5--9.0 to 10.5--11.0 dex, since bins outside this range contain too few galaxies to be statistically reliable.

The median size of the HST-Dark galaxies is smaller than that of the general population in every stellar mass bin, matching observations from the UNCOVER data \citepalias{Biswas2026}, indicating a fundamental difference between these galaxies and the typical galaxy population: HST-Dark galaxies have systematically higher mass densities. Several effects could contribute. First, a given amount of dust produces higher attenuation in a more compact galaxy with an optically thicker ISM (since in the SED fitting, dust is modeled as a uniform foreground screen, which is optically thicker for a smaller galaxy); this screens more starlight and pushes the source into the optically dark regime of our selection. Second, dust forms through baryonic interactions and stellar processes; a denser galaxy would have higher star formation, more frequent violent events (e.g., supernovae (SNe) and gamma ray bursts) and a higher rate of baryonic interactions, allowing rapid dust build-up and an optically dark nature. Third, galaxies with older stellar populations, redder starlight, and pre-existing dust would have a lower SNe rate and, hence, fewer energetic perturbations of the gravitational potential, allowing them to become denser. The HST-Dark galaxies are therefore systematically denser sources for several physical reasons. These hypotheses require further testing with simulations and deeper data.

%% file: conclusion.tex
To conclude, we have robustly selected and characterized the general physical properties of HST-Dark galaxies in $\sim 270$ arcmin$^2$ of the UDS field from the first data release of the JWST/MINERVA medium-band survey (UDS-DR0), combined with PRIMER and archival HST imaging. Beginning with a SUPER photometric catalog of 185,133 sources, we modeled the SEDs of the population with BAGPIPES utilizing photometry from up to 33 HST, NIRCam, and MIRI bands. Next, we selected HST-Dark candidates with $26 \leq \mathrm{F356W} \leq 27.5$ mag and $\mathrm{F150W} - \mathrm{F356W} \geq 1.5$ mag, removed spurious sources and artifacts through several rounds of careful visual inspection, and applied a goodness-of-fit cut of $\Delta z/(1+z) \leq 0.2$ to construct a final science sample of 372 HST-Dark galaxies, 105 of which have grade 3 NIRSpec spectra from the DAWN JWST Archive. We reran the SED modeling for sources with available spectra by fixing the redshift as the measured spectroscopic redshift, to reduce uncertainties in modeling. We also measured non-parametric morphological quantities for the full MINERVA catalog with \texttt{statmorph}. A summary of our most important inferences and conclusions is as follows.

\begin{enumerate}
    \item The redshift distribution is bimodal, with a clear minimum at $z = 2$; we adopt this as our completeness limit. Above $z>2$, our sample contains 325 HST-Dark galaxies with a sky surface density of $\sim 1.20$ arcmin$^{-2}$, a larger sample than any previous study on HST-Dark galaxies. The sky density of sources is also the highest of all previous studies except the lensed UNCOVER sample, which probes a fainter, lower-mass population. This allows us to derive representative statistical properties with much tighter constraints due to reduced Poisson fluctuations and cosmic variance.

    \item The $z > 2$ population, referred to as the final HST-Dark population, is comprised of high-redshift, massive, moderately star-forming and very dusty galaxies, with median $z = 4.15$, $\log(M_\star/M_\odot) = 9.64$, SFR $= 22.45\,M_\odot\,\mathrm{yr}^{-1}$, and $A_V = 1.85$. Thus, the MINERVA survey samples the median HST-Dark population, located between the extreme SMGs (starbursts) and Spitzer-selected galaxies at the high-mass end and the faint (lensed) UNCOVER galaxies at the low-mass end, with properties consistent with the much smaller, diagnostic CEERS sample.

    \item The 47 galaxies below completeness, with $z < 2$, are an incomplete sample of proximal (median $z \sim 0.62$), very low stellar mass (median $M_\star\sim 10^{7.82}\,M_\odot$) and extremely dusty (median $A_V \sim 3.42$) galaxies with minimal star formation. All but 4 of these galaxies satisfy the HELM criterion proposed by \citetalias{Bisigello2025} (Eqs.~\ref{Eq1:HELM_Criterion_1} \& \ref{Eq2:HELM_Criterion_2}).

    \item The HST-Dark galaxies have much higher dust attenuation $A_V$ and, on average, higher sSFR than the general MINERVA population. Within the complete, $z > 2$ sample, sSFR increases with redshift and $A_V$ shows a weak anticorrelation with redshift, since HST-darkness can arise from either redness or faintness, and higher redshift compensates for lower attenuation.

    \item Our galaxies follow the star-forming main sequence (SFMS) over $M_\star \sim10^{8.5}$--$10^{10.5}\,M_\odot$, with dust attenuation increasing towards the upper end of the sequence. However, their scatter about the SFMS exceeds the measurement uncertainties, the redshift evolution of the pre-JWST SFMS relation of \citetalias{Schreiber2015} and the SFMS scatter due to bursty star formation studied and quantified by \citet{Simmonds2025}, indicating that the empirical of the star-forming main sequence based on legacy imaging of HST-bright galaxies must be revisited with JWST samples in the near future, to more accurately trace the properties and evolution of the entire population of galaxies.

    \item In the rest-UVJ colors diagram, nearly all our HST-Dark sources are star-forming, divided almost equally between dusty and non-dusty classes. The non-dusty fraction increases with redshift. We find a very small fraction ($\sim 2.5\%$) of HST-Dark quiescent galaxies, in contrast to previous studies of HST-Dark galaxies at similar redshifts by \citetalias{PerezGonzalez2023} \& \citetalias{Barrufet2025}, but in agreement with the $\sim 3\%$ fraction found by \citetalias{Biswas2026}. The large sample sizes of the MINERVA and UNCOVER populations suggest that the high quiescent fraction of HST-Dark galaxies proposed by \citetalias{PerezGonzalez2023} and \citetalias{Barrufet2025} ($\sim 13-18\%$) might be overestimated, and a fraction of $\lesssim 3\%$ may be more statistically significant. Follow-up studies analyzing the available NIRSpec spectra of these quiescent sources would better characterize their nature, and also hint to reasons for their measured abundance.

    \item The contribution of HST-Dark galaxies (upper error bars) to the redshift evolution of the CSFRD reaches the upper limit of the dust-obscured CSFRD model proposed by \citetalias{Zavala2021} at $z \sim 6.5$, approaching $\sim 30\%$ of the dust-unobscured CSFRD estimated by \citet{Bouwens2022} in the highest-redshift bins ($7<z<8$). Because of the flux limits of our survey, we do not detect the entire population of HST-Dark galaxies, predominantly missing the low stellar mass component of this population. Hence, these estimates are lower limits, and a substantial part of the obscured cosmic star formation at early epochs may still be missing from current censes; this motivates the need to re-inspect the plot of cosmic star formation history first reported by \citet{Madau2014}, with a thorough account of these hidden sources from deeper observations.

    \item At $3 < z < 5$, HST-Dark galaxies have systematically smaller R$_{80}$ than the general MINERVA population across the typical range of stellar masses, implying higher stellar mass densities. Compact galaxies attenuate their UV/optical starlight more efficiently for the same given dust mass, and may build up dust faster through more frequent baryonic interactions, hence the HST-Dark sources are intrinsically denser than UV-bright galaxies.

    \item $\sim 52\%$ of our sample has medium-band coverage from MINERVA, relative to the $\sim 50\%$ of the UDS-DR0 area for which medium band imaging has been released; hence, the shallower MINERVA medium bands do not appear to increase the number of selected sources relative to the deep PRIMER broad bands alone.
\end{enumerate}

These results demonstrate that a wide-area, multiband JWST survey can characterize the median HST-Dark population at high redshifts in a statistically significant manner. These galaxies seem to be substantial contributors to global cosmological relationships like the integrated cosmic star formation history at $z > 4$. The full MINERVA survey, covering $\sim 542$ arcmin$^2$ across four CANDELS fields, will test these conclusions against cosmic variance and extend them to much lower masses. Follow-up high-quality NIRSpec spectroscopy must also be incorporated into such analyses, making use of imaging studies like ours to detect targets for future spectroscopic surveys. A combination of NIRSpec spectra and a more robust goodness-of-fit criterion will be essential to resolve the remaining degeneracies in modeling, to establish the true quiescent fraction of this population, to recalibrate the main sequence and the cosmic star formation history; and ultimately, to robustly bring these long-hidden, elusive astronomical sources from the cosmic noon into the light.